\documentclass[12pt,preprint,iop]{emulateapj}
\usepackage{amsfonts,amsmath,apjfonts}
\usepackage{natbib}
\usepackage{graphicx}
\usepackage[caption=false]{subfig}
\usepackage{booktabs}
\usepackage[breaklinks,colorlinks,citecolor=blue]{hyperref}
\usepackage[all]{hypcap}

\def\cfa{1}

\shorttitle{PS16dtm: A Tidal Disruption Event in a NLS1 Galaxy}
\shortauthors{Blanchard et al.}

\begin{document}

\title{PS16dtm:  A Tidal Disruption Event in a Narrow-line Seyfert 1 Galaxy}

\author{
P. K. Blanchard\altaffilmark{\cfa}, 
M. Nicholl\altaffilmark{\cfa},
E. Berger\altaffilmark{\cfa},
J.~Guillochon\altaffilmark{\cfa},
R.~Margutti\altaffilmark{2},
R.~Chornock\altaffilmark{3}, 
K.~D.~Alexander\altaffilmark{\cfa},
J.~Leja\altaffilmark{\cfa},
M.~R.~Drout\altaffilmark{4,5}
}
\email{pblanchard@cfa.harvard.edu}

\altaffiltext{1}{Harvard-Smithsonian Center for Astrophysics, 60 Garden Street, Cambridge, MA 02138, USA}
\altaffiltext{2}{Center for Interdisciplinary Exploration and Research in Astrophysics (CIERA) and Department of Physics and Astronomy, Northwestern University, Evanston, IL 60208}
\altaffiltext{3}{Astrophysical Institute, Department of Physics and Astronomy, 251B
Clippinger Lab, Ohio University, Athens, OH 45701, USA}
\altaffiltext{4}{Carnegie Observatories, 813 Santa Barbara Street, Pasadena, CA
91101, USA 2}
\altaffiltext{5}{Hubble, Carnegie-Dunlap Fellow}

\begin{abstract}
\medskip

We present multi-wavelength observations of PS16dtm (also known as SN\,2016ezh), an optically discovered super-luminous transient that occurred at the nucleus of SDSSJ015804.75-005221.8, a known Narrow-line Seyfert 1 galaxy hosting a $\sim$10$^6$ M$_\odot$ black hole.  The transient was previously claimed to be a Type IIn SLSN due to its luminosity and hydrogen emission lines.  The light curve shows that PS16dtm brightened by about two magnitudes in $\sim$50 days relative to the archival host brightness and then exhibited a plateau phase for about $\sim$100 days followed by the onset of fading in the UV.  During the plateau PS16dtm showed no color evolution, maintained a steady blackbody temperature of $\sim$1.7 $\times 10^{4}$ K, and radiated at approximately the Eddington luminosity of the supermassive black hole.  The spectra, spanning UV to near-IR, exhibit multi-component hydrogen emission lines and strong \ion{Fe}{2} emission complexes, show little evolution with time, and closely resemble the spectra of NLS1 galaxies while being distinct from those of Type IIn SNe.  In addition, PS16dtm is undetected in the X-rays by \textit{Swift}/XRT to a limit an order of magnitude below an archival \textit{XMM-Newton} detection of its host galaxy.  These observations strongly link PS16dtm to activity associated with the supermassive black hole and are difficult to reconcile with a SN origin.  Moreover, the properties of PS16dtm are unlike any known form of AGN variability, and therefore we argue that it is a tidal disruption event in which the accretion of the stellar debris powers the rise in the continuum and excitation of the pre-existing broad line region, while at the same time providing material that obscures the X-ray emitting region of the pre-existing AGN accretion disk.  A detailed TDE model fits the bolometric light curve and indicates that PS16dtm will remain bright for several years; we further predict that the X-ray emission will reappear on a similar timescale as the accretion rate declines.  Finally, we place PS16dtm in the context of other TDEs and find that tidal disruptions in active galaxies are an order of magnitude more efficient and reach Eddington luminosities, likely due to interaction of the stellar debris with the pre-existing accretion disk.                

\smallskip
\end{abstract}


\section{Introduction}

The proliferation of wide-field untargeted optical time-domain surveys over the past decade has enabled the discovery of a diverse set of observationally and volumetrically rare luminous transients.  These include superluminous supernovae \citep[SLSNe;][]{Chomiuk2011,Quimby2011,GalYam2012,Nicholl2015} with hydrogen (Type II) and without hydrogen (Type I) and tidal disruption events \citep[TDEs;][]{Gezari2009,Velzen2011,Chornock2014,Komossa2015}, as well as some events that defy a clear classification \citep[e.g. ASASSN-15lh;][]{Dong2016,Raf2016,Leloudas2016}.  Among these various events there are some transients that coincide with the nuclei of host galaxies, in some cases with clear signatures of active galactic nuclei (AGN).  In such cases the nature of the transients has been mostly ambiguous, with possible interpretations spanning nuclear supernovae, TDEs, and unusual AGN flaring activity.  

A well-studied example is the luminous ($M_V\approx -22.7$ mag) transient CSS100217:102913+404220 \citep[hereafter CSS100217;][]{Drake2011}, discovered by the Catalina Real-time Transient Survey \citep[CRTS;][]{Drake2009} in coincidence with the nucleus of a Narrow-line Seyfert 1 galaxy (NLS1).  The transient was characterized by a brightening of about 1.2 mag relative to the quiescent galaxy brightness within about 80 days, followed by a fading back to the baseline level in $\sim 1$ year.  The spectrum of the transient was similar to the quiescent galaxy spectrum, with bright multi-component hydrogen Balmer lines, and notable \ion{Fe}{2} lines.  \citet{Drake2011} explored an origin of this event as an AGN flare, TDE, or a Type IIn SLSN, and argued for the latter primarily because the level of variability was unprecedented for NLS1 galaxies.  They argued the temperature and decline rate were inconsistent with theoretical predictions for TDEs.   

Here we present radio, near-IR, optical, UV, and X-ray observations of the recent transient PS16dtm at $z = 0.0804$, which has been claimed as a Type IIn SLSN \citep{asiagoATel,DongATel}, as well as archival multi-wavelength observations of its host galaxy.  We show that the host is a well-studied NLS1 galaxy, that PS16dtm coincides with its nucleus, and that the spectrum of the transient (dominated by multi-component hydrogen Balmer lines and strong \ion{Fe}{2} lines) closely resembles those of NLS1 galaxies, while being distinct from Type IIn SNe.  The UV/optical light curves are also highly unusual for a Type IIn SN.  In addition, we find that PS16dtm radiated with a peak luminosity equal to the Eddington luminosity of the supermassive black hole at the center of the NLS1 host galaxy.  Surprisingly, we find about an order of magnitude drop in the X-ray emission during the transient outburst relative to an archival X-ray detection of the NLS1 host galaxy, suggesting a link between PS16dtm and the nucleus of its host.  Considering both extreme AGN variability and a TDE interpretation, we argue that only a TDE can explain all of the observed properties of PS16dtm.  The similarity of CSS100217 and PS16dtm suggests that CSS100217 may also be a TDE.    

In Section 2 we present the properties of the host galaxy of PS16dtm.  In Section 3 we present our own and public observations of PS16dtm.  In Section 4 we discuss the light curve, spectra, and X-ray properties.  In Section 5 we compare the spectra of PS16dtm to Type IIn SNe, NLS1 galaxies, and TDEs.  In Section 6 we summarize the properties of PS16dtm and argue that a Type IIn SN or AGN variability cannot explain all of these properties.  In Section 7 we present a discussion of why a TDE provides the best explanation of PS16dtm.  In Section 8 we discuss PS16dtm in the broader context of TDEs and we conclude in Section 9.   

In this paper we use $H_{0} = 67$ km s$^{-1}$ Mpc$^{-1}$, $\Omega_{m} = 0.32$, and $\Omega_{\Lambda} = 0.68$ \citep{Planck2013}, resulting in a luminosity distance of 381 Mpc to PS16dtm.  The Galactic extinction along the sight line to PS16dtm is $E(B-V) = 0.0222 \pm 0.0004$ \citep{SF2011}.  Sloan Digital Sky Survey (SDSS) spectra presented in this paper were obtained from Data Release 13 \citep{SDSS2016}.

\begin{figure*}[t!]
\begin{center}
\subfloat[]{\includegraphics[scale=0.35]{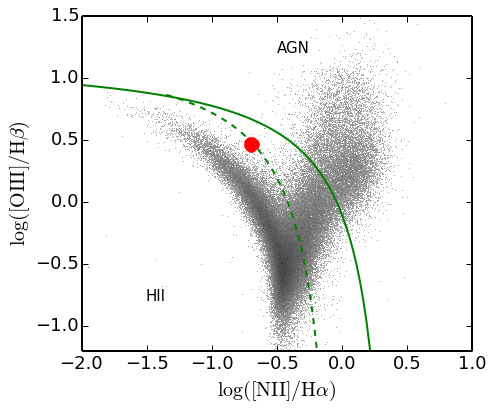}}
\subfloat[]{\includegraphics[scale=0.35]{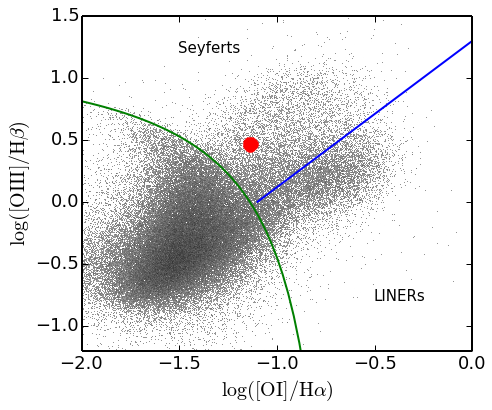}}
\subfloat[]{\includegraphics[scale=0.35]{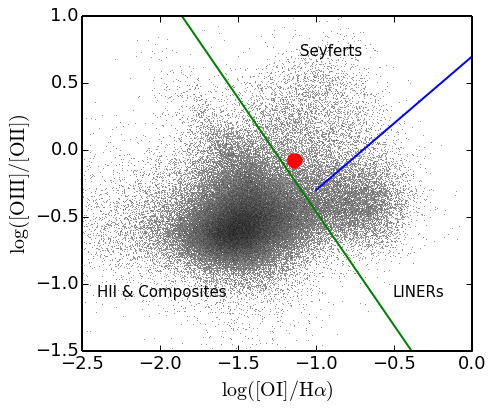}}
\end{center} 
\begin{center}
\subfloat[]{\includegraphics[scale=0.45]{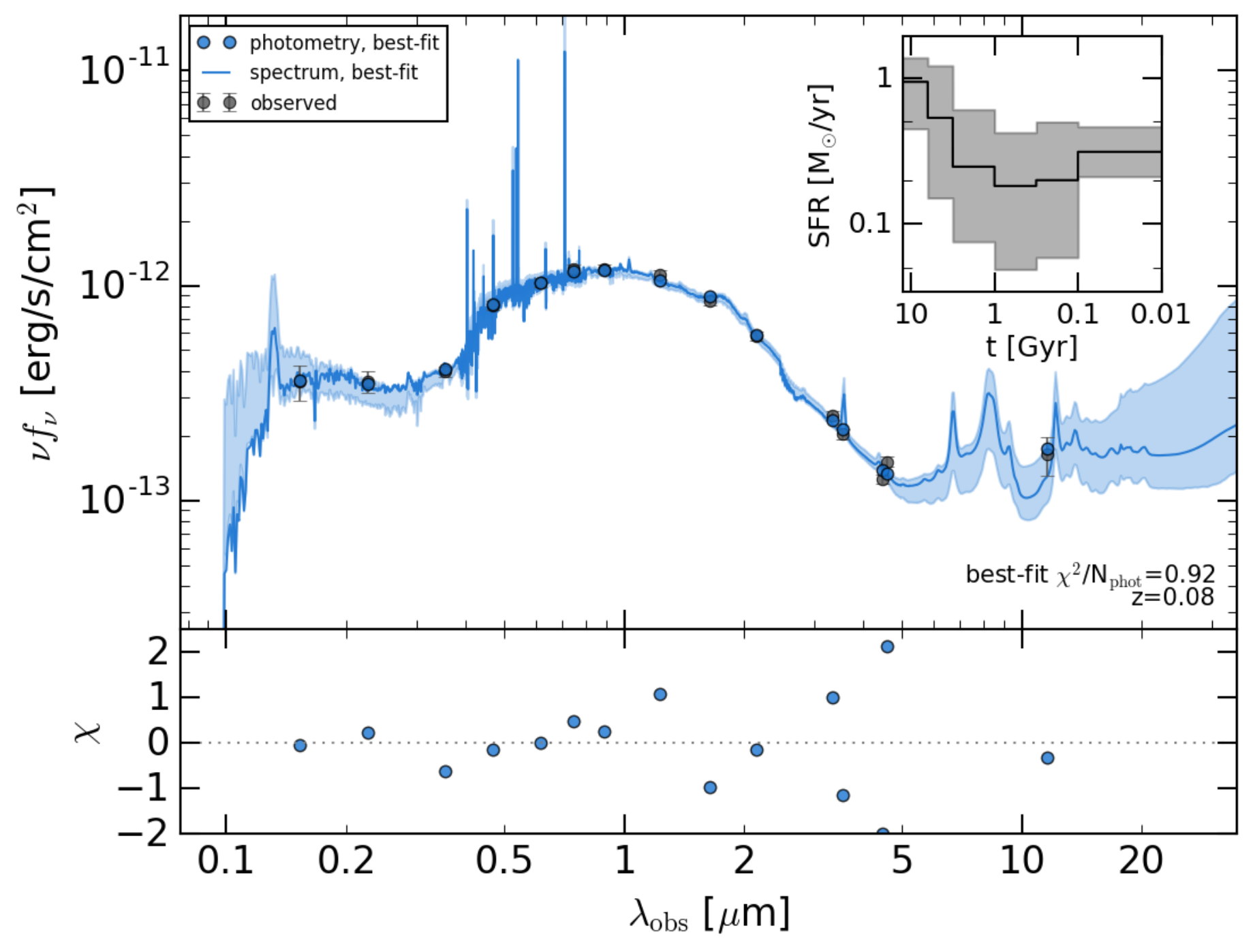}}
\subfloat[]{\includegraphics[scale=0.315]{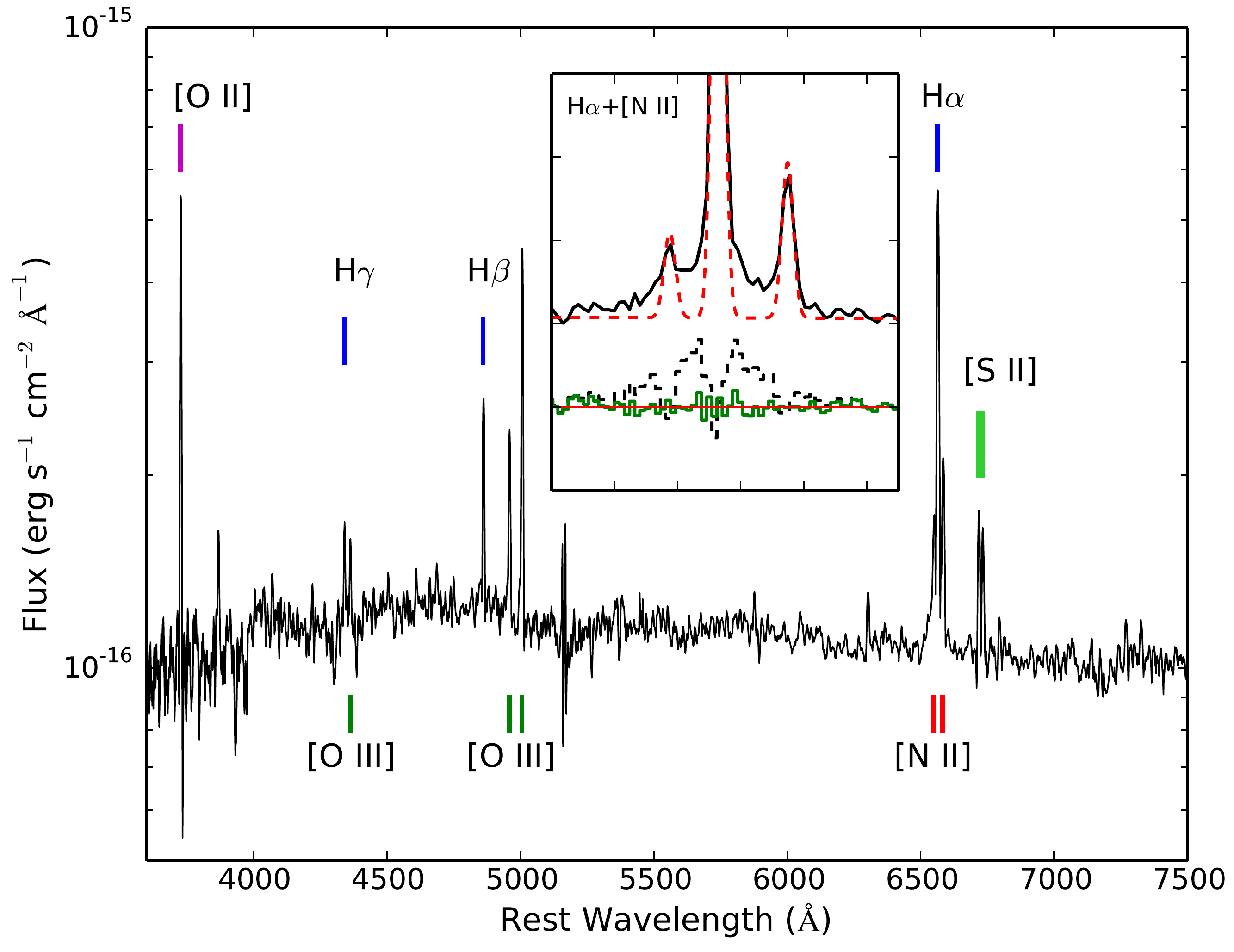}}
\end{center}
\caption{Emission line ratios, spectral energy distribution, and SDSS spectrum of SDSSJ0158.75-005221.8, the host galaxy of PS16dtm.  Panels (a), (b), and (c) show various line ratio diagnostic diagrams involving the line ratios [OIII]/H$\beta$, [NII]/H$\alpha$, [OI]/H$\alpha$, and [OIII]/[OII] \citep{BPT,Kewley2006}.  The red dot marks the position of PS16dtm's host galaxy.  The shaded regions are 2D histograms representing the locations of SDSS galaxies with $0.04 < z < 0.1$ measured using line fluxes from the MPA-JHU data release\footnote{\url{http://wwwmpa.mpa-garching.mpg.de/SDSS/DR7/}}.  Darker regions indicate a higher density of galaxies.  The extreme star formation classification line \citep[solid green;][]{Kewley2001}, the pure star formation line \citep[dashed green;][]{Kauffmann2003}, and the Seyfert-LINER classification line \citep[blue;][]{Kewley2006} are shown.  (d) The best-fit photometry and spectrum of the Prospector-$\alpha$ model (blue) are compared to the observed photometry (grey). The 16th and 84th percentiles of the model spectrum are shaded in light blue. The 16th, 50th, and 84th percentiles of the marginalized star formation history are in the inset panel. The residuals between the observations and the best-fit photometry are shown in the lower panel.  (e) The SDSS spectrum of the host galaxy with an inset zoomed in on H$\alpha$.  The inset shows that there are significant residuals when fitting H$\alpha$ with only a narrow component fixed at the instrumental resolution (dashed red for fit, dashed black for residuals), indicating the presence of an additional broad component, a signature of a Seyfert 1 AGN.  The residuals resulting from including a broad ($\sim1680$ km s$^{-1}$) component are also shown (solid green).}
\label{host}
\end{figure*}

\section{The Narrow-Line Seyfert 1 Host of PS16dtm}
\label{sec:host}

\begin{figure}[ht!]
\begin{center}
\includegraphics[scale=0.4]{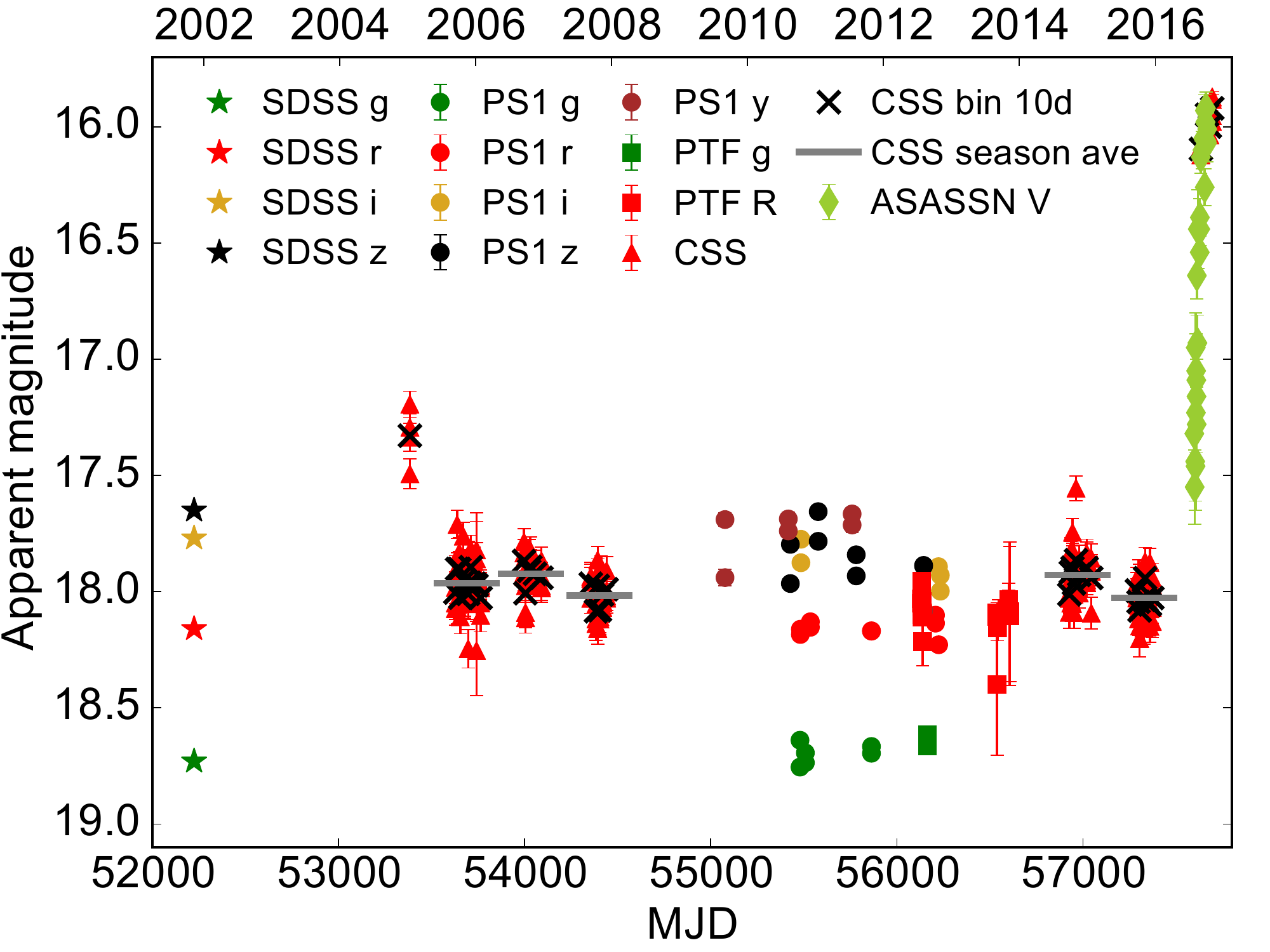}
\end{center}
\caption{The historic light curve of PS16dtm's host galaxy since 2002, including Catalina Sky Survey (CSS), SDSS, Palomar Transient Factory (PTF), and Pan-STARRS (PS1) data.  There is evidence of a possible brightening episode at the first epoch of CSS observations and modest variability between CSS seasonal averages.}
\label{historicLC}
\end{figure}

The pre-outburst properties of SDSSJ015804.75-005221.8, the host galaxy of PS16dtm, indicate
that it is a narrow-line Seyfert 1 \citep{op85}.  Emission
line ratios from an archival SDSS spectrum show a clear separation
from the locus of star forming galaxies and overlap with Seyfert
galaxies (Figure~\ref{host}).  Moreover, the host galaxy was
included in the Seyfert 1 samples of \citet{gh04} and \citet{gh07},
which were selected based on a broad component to H$\alpha$ in SDSS
spectra (see Figure~\ref{host}).  Follow-up spectroscopy by
\citet{xbg+11} using the Magellan Echellete spectrograph indicates
FWHM(H$\alpha)\approx 1200$ km s$^{-1}$. These authors also
measure a stellar velocity dispersion of $\sigma_*\approx 45$ km
s$^{-1}$, from which they infer a black hole mass of $M_{\rm
BH}\approx 10^{5.9}$ M$_\odot$.  Using the SDSS spectrum and the relation from \citet{Bentz2013} between AGN luminosity and radius of the broad line region (assuming the velocity width of H$\alpha$ corresponds to a circular Keplerian velocity)  
we find a similar, though slightly higher, black hole mass and therefore use 10$^6$ M$_\odot$ in subsequent analysis.
Analysis of {\it Hubble Space Telescope} Wide-Field Planetary Camera 2 (WFPC2) observations by \citet{jgh+11} indicates a
bulge-dominated morphology (Sersic index of $3$) with an effective
radius of about $1.6$ kpc, and an estimated AGN flux contribution in
the F814W filter of about 5\% based on radial profile fitting.  In Figure \ref{historicLC} we show 
the historical optical light curve of the galaxy which exhibits only mild variability 
prior to PS16dtm.

In addition, the galaxy was observed and detected with \textit{XMM-Newton} in July 2005
(Program 030311; PI: Ptak).
A spectral analysis by \cite{Pons14} in the $0.3-10$ keV range indicates a best fitting power-law model with photon index $\Gamma=2.2\pm0.2$ (90\% confidence level), and no statistically significant evidence for intrinsic neutral hydrogen absorption, or a soft excess.  The inferred luminosity is $L_{\rm X}=(1.2\pm0.5)\times10^{42}\,\rm{erg\,s^{-1}}$ \citep{Pons14}, 
typical of NLS1 galaxies \citep[e.g.,][]{gwl+04}.  The
X-ray luminosity corresponds to $\sim 0.01\,L_{\rm Edd}$ for the
inferred black hole mass, also typical of NLS1 galaxies (e.g.,
\citealt{bg92}). The galaxy is not detected in the FIRST radio survey,
nor in our deeper follow-up VLA observations (see Section \ref{sec:radio}) which place a $3\sigma$
upper limit of $\lesssim 24$ $\mu$Jy at 6 GHz, corresponding to
$\nu$$L_{\nu,r}\lesssim 2.3\times 10^{37}$ erg s$^{-1}$.  The ratio of
radio to X-ray luminosity, $R_X\lesssim -5.2$ is again typical of NLS1
galaxies \citep{tw03}.  Taking into account the inferred black hole
mass, this ratio is also consistent with the ``fundamental plane'' of
black hole activity (e.g., \citealt{mhm03}).

  Using archival photometry of the galaxy spanning from the UV to the
  IR (GALEX, SDSS, VISTA Hemisphere Survey, \textit{Spitzer}, WISE) we model
  the spectral energy distribution of the host galaxy using the {\tt
    prospector} software package \citep{ljc+16}; see
  Figure~\ref{host}.  We find a stellar mass of log($M$/M$_\odot$) = $9.94^{+0.06}_{-0.08}$, a metallicity of log($Z$/Z$_\odot$) = $-0.46^{+0.20}_{-0.30}$, a present-day
  star formation rate of 0.62$^{+0.28}_{-0.21}$ M$_\odot$ yr$^{-1}$, a declining
  star formation history, with a specific star formation rate of about
  0.07$^{+0.04}_{-0.03}$ Gyr$^{-1}$, and evidence\footnote{The modeling
    of the AGN component within {\tt prospector} will be described in
    an upcoming publication (Leja et al.~in prep.)} for AGN emission in the mid-infrared, with a
  bolometric luminosity contribution of about $5\%$, in agreement with the value
  inferred from modeling of the radial light profile in the \textit{HST} data
  \citep{jgh+11}.  Using this AGN contribution we calculate an optical to X-ray luminosity 
  ratio, $L_{\rm opt}/L_{\rm X} \sim 0.5$, using $\lambda L_{\lambda}$(5100\AA) to estimate 
  the optical luminosity from the archival SDSS spectrum and $L_{\rm X}$ from the \textit{XMM-Newton} 
  detection above.  The ratio we find is consistent with typical values for NLS1 galaxies \citep{gwl+04}.

\begin{figure*}[ht!]
\begin{center}
\includegraphics[width=6in]{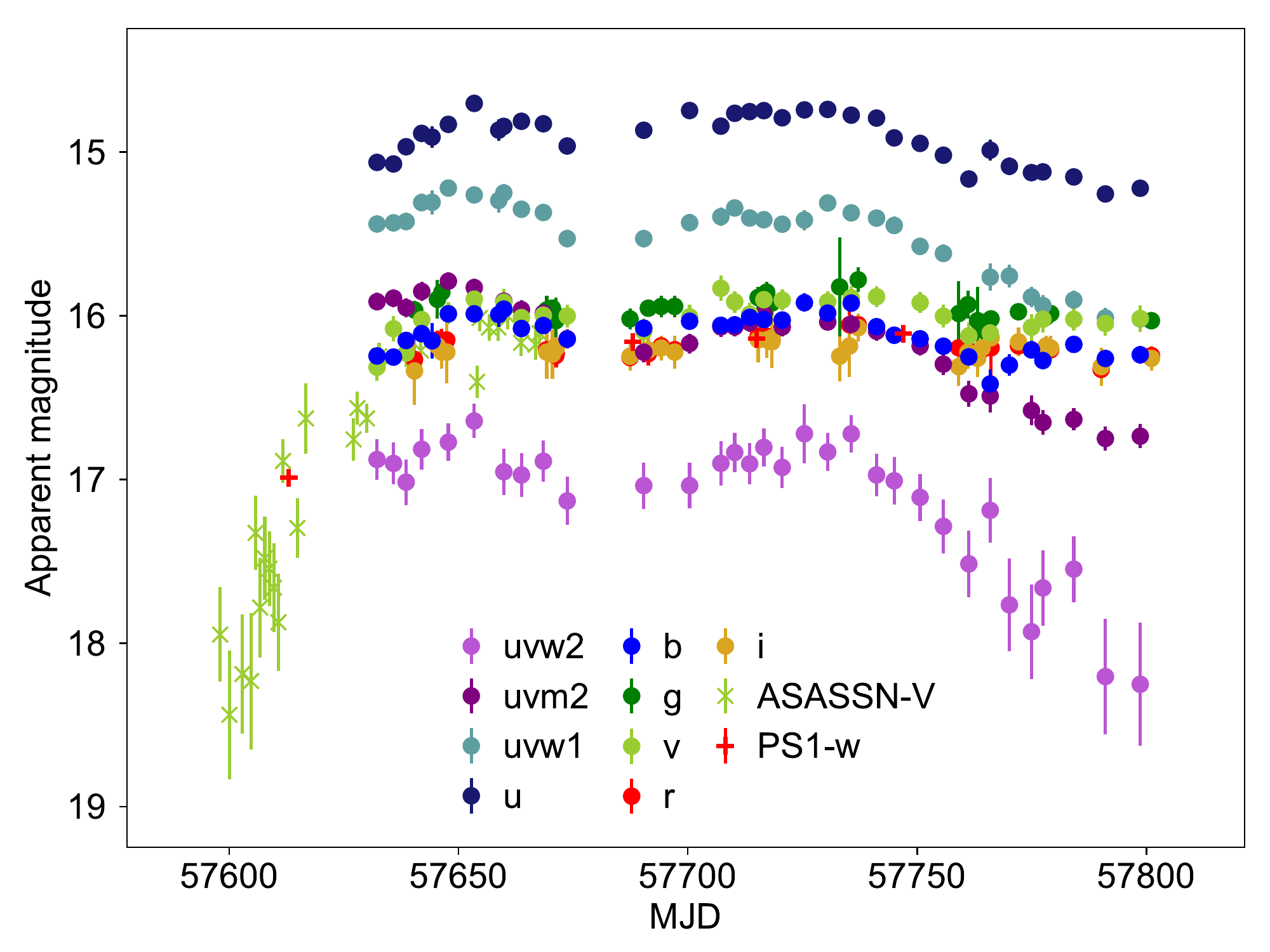}
\end{center}
\caption{Host-subtracted and Galactic extinction-corrected UV/optical light curves of PS16dtm including our own $gri$ observations, PS1 $w$-band, archival $\textit{Swift}$/UVOT data, and $V$-band data reported by ASASSN \citep{DongATel}.}
\label{LC}
\end{figure*}

\begin{figure}[ht!]
\begin{center}
\vspace{3mm}
\includegraphics[scale=0.33]{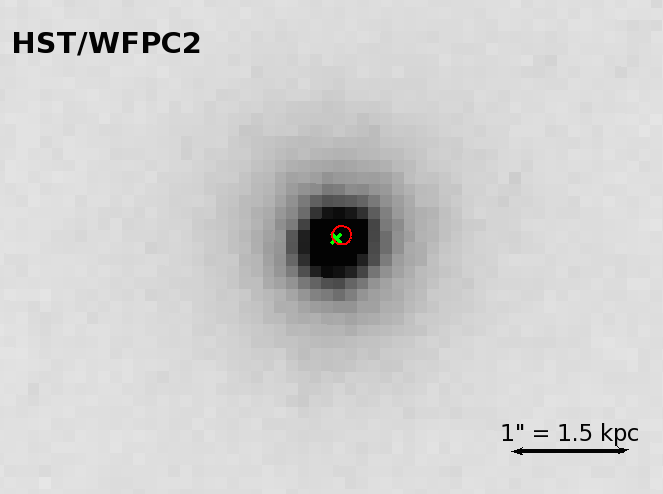}
\end{center}
\caption{Archival \textit{HST} image of SDSSJ015804.75-005221.8 showing the position of the galaxy centroid (green "x") and a $1\sigma$ error circle (red) centered on the position of PS16dtm as determined by relative astrometry using our ground-based images.  North is up and East is to the left.  The offset is $0.03\arcsec \pm 0.05\arcsec$, indicating that PS16dtm is consistent with being located at the nucleus of SDSSJ015804.75-005221.8.}
\label{astrometry}
\end{figure}

\section{Observations of PS16dtm}

PS16dtm was discovered by the Pan-STARRS Survey for Transients \citep[PSST;][]{Huber2015} on 12 August 2016 with a $w$-band AB magnitude of 17.1 \citep{PSSTATel}, and was classified as a Type II SN on 26 August 2016 based on spectroscopic observations that showed hydrogen Balmer emission lines \citep{asiagoATel}.  PS16dtm was also independently discovered by CRTS and designated CSS160902-015805-005222 and was detected by the All-Sky Automated Survey for Supernovae \citep[ASAS-SN;][]{DongATel}.  The redshift of the apparent host galaxy SDSSJ015804.75-005221.8 ($z$ = 0.0804), whose properties were discussed in Section \ref{sec:host}, implied an absolute magnitude of $-20.6$ for the PSST detection (later reaching $-21.7$ in the ASAS-SN data), which led \citet{DongATel} to claim that this event is a Type IIn SLSN.  Examining the initial spectrum we noticed that it also exhibited a strong broad emission feature at $5000 - 5400$ \AA\ due to \ion{Fe}{2} multiplets, a feature that is seen in some supernovae but that is also often very prominent in NLS1 galaxies.  This observation combined with PSST astrometry suggesting a nuclear event motivated us to initiate follow-up observations.

\subsection{UV and Optical Photometry}
We obtained observations of PS16dtm in the $gri$ bands using the 48-inch telescope at the Fred Lawrence Whipple Observatory (FLWO) on Mt. Hopkins, Arizona.  We analyzed the data using standard techniques in IRAF and performed aperture photometry on PS16dtm using an aperture encompassing the transient and all of the galaxy light.  Calibrated magnitudes were obtained on the AB system using reference stars from the Pan-STARRS 3$\pi$ catalog.     

We also analyzed \emph{Swift} UV Optical Telescope \citep[UVOT;][]{Roming05} observations obtained between 1 September 2016 and 14 February 2017. We followed the prescriptions of \cite{Brown09} with the updated calibration files and revised zero points of \cite{Breeveld11}.  Calibrated UVOT magnitudes in the $uvw2$, $uvm2$, $uvw1$, $u$, $b$, and $v$ filters are reported in Vega magnitudes.    

To measure the flux of PS16dtm only, we subtracted archival flux measurements of the host galaxy for each optical and UV band, using the SED presented in Section \ref{sec:host} to estimate host fluxes in the UVOT bands.  The resulting host-subtracted light curves are shown in Figure \ref{LC}, corrected for Galactic extinction.  We also plot the rise portion of the light curve reported by the ASAS-SN team in the $V$ band \citep{DongATel}, including the same extinction and host corrections.   

\subsection{Astrometry}
To locate PS16dtm within its host galaxy we perform relative astrometry between our best-seeing (${\rm FWHM} \approx 1.8"$) FLWO 48-inch images and an archival $\textit{HST}$/WFPC2 image of the host galaxy.  To measure the centroid of the transient we subtract the host galaxy contribution from the 48-inch using a pre-outburst image from the Pan-STARRS 3$\pi$ Survey and the {\tt HOTPANTS} image subtraction software\footnote{\url{http://www.astro.washington.edu/users/becker/v2.0/hotpants.html}}.  Using 4 stars in common between the 48-inch and $\textit{HST}$ images we match the images and find that PS16dtm is offset by 0.05\arcsec\,$\pm$ 0.08\arcsec\ ($76 \pm 122$ pc) from the center of its host galaxy; the uncertainty is dominated by the astrometric tie, with minor contribution from uncertainty in the centroids of the bright transient and host galaxy.  We show the position of PS16dtm on the \textit{HST} image in Figure \ref{astrometry}.  We find a similar result (0.03\arcsec\,$\pm$ 0.05\arcsec, or $45 \pm 76$ pc) by using the archival 3$\pi$ image to measure the host centroid, utilizing 30 stars in common between the two images.  PS16dtm is therefore consistent with having occurred at the nucleus of its host galaxy.

To further assess the location of PS16dtm, we also obtained \textit{HST} imaging of PS16dtm through an \textit{HST} Director's Discretionary Time program (14902; PI: Blanchard).  Using the Space Telescope Imaging Spectrograph (STIS) with the CCD detector and 50CCD (Clear) filter, we obtained four dithered images that we drizzle-combined using {\tt astrodrizzle}\footnote{\url{http://drizzlepac.stsci.edu}} to improve the spatial resolution ({\tt final\_scale} = 0.025$\arcsec$/pix, {\tt final\_pixfrac} = 0.8).  We clearly detect PS16dtm as a point source with no evidence for extended emission from the host galaxy.  While there are not enough sources in this image to perform relative astrometry, the lack of an asymmetry in the light distribution indicates that PS16dtm occurred at a small offset.   

\subsection{UV, Optical, and NIR Spectroscopy}
We obtained 8 epochs of optical spectroscopy using the FAST Spectrograph \citep{Fabricant1998} on the 60-inch telescope at FLWO, the Blue Channel Spectrograph \citep{Schmidt1989} on the 6.5-m MMT, the Ohio State Multiple Object Spectrograph \citep[OSMOS;][]{Martini2011} on the 2.4-m Hiltner telescope at MDM Observatory, the Wide-Field Reimaging CCD (WFCCD) on the 100-inch DuPont telescope, and the Magellan Echellete Spectrograph \citep[MagE;][]{Marshall2008} on the 6.5-m Magellan Baade telescope.  The FAST spectra were obtained on 5 September 2016, 1 October 2016, and 1 November 2016 using the 300 gpm grating, a 3$\arcsec$\ slit, and covered the range $3200 - 7000$ \AA\ in the rest-frame of PS16dtm with a resolution of $\sim$5.7\AA.  The Blue Channel spectrum was obtained on 5 December 2016 using the 300 gpm grating, a 1$\arcsec$\ slit, and covered the rest-frame range $3000 - 8000$ \AA\ with a resolution of $\sim4$\AA.  The OSMOS spectra were obtained on 15 December 2016 and 11 February 2017 using the 1.2$\arcsec$\ center slit with the VPH-red grism and an OG530 order-blocking filter.  The spectrum covered the rest-frame range 4900--9300~\AA\ with a resolution of $\sim$5\AA.  The MagE spectrum was obtained on 28 January 2017 using a 1$\arcsec$\ slit, covering the rest-frame range $3000 - 9200$ \AA\ with a resolution of $\sim$1\AA.  The WFCCD spectrum was obtained on 19 February 2017 using the blue grism with a 1.65$\arcsec$\ slit, yielding a resolution of $\sim$8\AA.  We analyzed most spectra using standard techniques in IRAF to obtain 1D wavelength- and flux-calibrated spectra.  The MagE spectrum was reduced using a specialized reduction pipeline provided by the Carnegie Observatories Software Repository\footnote{\url{http://code.obs.carnegiescience.edu}}.  Relative flux calibrations were applied using standard stars observed on the same night.  The absolute flux calibrations were checked against photometry taken at similar epochs.  The spectra were then corrected for Galactic extinction and transformed to the rest-frame of PS16dtm for analysis.        

We obtained one epoch of near-infrared (NIR) spectroscopy on 5 February 2017 using the Folded-port Infrared Echellete \citep[FIRE;][]{Simcoe2013} on the 6.5-m Magellan Baade telescope.  We used the low-dispersion mode with a 1$\arcsec$\ slit to obtain a spectrum covering the rest-frame range $0.75 - 2.3$  $\mu m$ with a resolution of $\sim 24$\AA.  The FIRE spectrum was reduced using the {\tt FIREHOSE} reduction pipeline \citep{Simcoe2013}.  

We also obtained a near-ultraviolet (NUV) spectrum of PS16dtm through our \textit{HST} Director's Discretionary Time program.  The spectrum was obtained on 22 February 2017 using STIS with the NUV-MAMA detector, G230L grating, and 0.2$\arcsec$\ slit.  This yielded a spectrum covering the rest-frame range $1500 - 2900$ \AA\ with a resolution of $\sim$4\AA. 

The spectroscopic observations are summarized in Table \ref{tab:spec}.

\capstartfalse
\begin{deluxetable*}{ccccccc}[h!]
\tablecolumns{7}
\tabcolsep0.1in\footnotesize
\tablewidth{0pc}
\tablecaption{Spectroscopic Observations of PS16dtm   
\label{tab:spec}}
\tablehead {
\colhead {Date}   &
\colhead {MJD}     &
\colhead {Phase\tablenotemark{a}} &
\colhead {Telescope} &
\colhead{Instrument}  &
\colhead {Airmass}   &
\colhead {Resolution (\AA)}           
}   
\startdata
5 September 2016 & 57637.3 & +35  & FLWO 60-inch & FAST & 1.3 & 5.7 \\
1 October 2016 & 57663.4 & +59  & FLWO 60-inch & FAST & 1.2 & 5.7 \\
1 November 2016 & 57694.2 & +87  & FLWO 60-inch & FAST & 1.2 & 5.7 \\
5 December 2016 & 57728.8 & +119  & MMT & Blue Channel & 1.5 & 4 \\
15 December 2016 & 57737.5 & +127  & MDM & OSMOS & 1.2 &  5 \\
28 January 2017 & 57782.5 & +169 & Magellan/Baade & MagE & 1.7 & 1 \\
5 February 2017 & 57790.5 & +176 & Magellan/Baade & FIRE & 1.9 & 24 \\
11 February 2017 & 57795.5 & +181 & MDM & OSMOS & 2.0 & 5 \\
19 February 2017 & 57804.5 & +189 & DuPont & WFCCD & 2.5\tablenotemark{b} & 8 \\ 
22 February 2017 & 57806.7 & + 191 & \textit{HST} & STIS/NUV-MAMA &  & 4  
\enddata
\tablenotetext{a}{Rest-frame days since MJD 57600}
\tablenotetext{b}{The high airmass of this observation affected the shape of the spectrum at the bluest wavelengths.}
\end{deluxetable*}
\capstarttrue

\subsection{X-Ray Observations}
Observations with the \emph{Swift} X-Ray Telescope (XRT; \citealt{Burrows05}) were obtained in conjunction with the UVOT data. We analyzed the XRT data using {\tt HEASOFT} (v6.19) and corresponding calibration files, and applied standard filtering and screening criteria (see \cite{Margutti13} for details). Using 52.9 ks of data collected between 1 September 2016 and 14 February 2017, we find no evidence for X-ray emission at the location of PS16dtm with a $3\sigma$ upper limit of $2.7\times10^{-4}$ $\rm{c\,s^{-1}}$ ($0.3-10$ keV).  We similarly find no detectable emission in the individual observations to a typical $3\sigma$ limit of $7.1\times10^{-3}$ $\rm{c\,s^{-1}}$.  The Galactic neutral hydrogen column density in the direction of PS16dtm is $N_{\rm H,MW}=2.5\times10^{20}$ $\rm{cm^{-2}}$ \citep{Kalberla05}.

Using the spectral parameters determined from the archival \textit{XMM-Newton} detection discussed in Section \ref{sec:host}, the XRT count-rate translates to an unabsorbed flux limit,  $F_X<1.0\times 10^{-14}$ $\rm{erg\,s^{-1}\,cm^{-2}}$ ($0.3-10$ keV), or $L_X<1.7\times 10^{41}$ $\rm{erg\,s^{-1}}$.  This is an order of magnitude lower than the archival \textit{XMM-Newton} detection.

\subsection{Radio Observations}
\label{sec:radio}
We observed PS16dtm using the Karl G. Jansky Very Large Array (VLA) on 22 September 2016 and 21 December 2016. We analyzed the data using the Common Astronomy Software Applications ({\tt CASA}) with 3C48 as a flux calibrator and J0215-0222 as a gain calibrator. We do not detect PS16dtm in either C band or K band on both epochs, resulting in $3\sigma$ upper limits of 23 $\mu$Jy at 6.0 GHz and 57 $\mu$Jy at 21.8 GHz on 22 September 2016 and 25 $\mu$Jy at 6.0 GHz and 51 $\mu$Jy at 21.8 GHz on 21 December 2016, measured using the {\tt imtool} program within the {\tt pwkit} package\footnote{\url{https://github.com/pkgw/pwkit}}. 

\begin{figure}[t!]
\begin{center}
\includegraphics[scale=0.37]{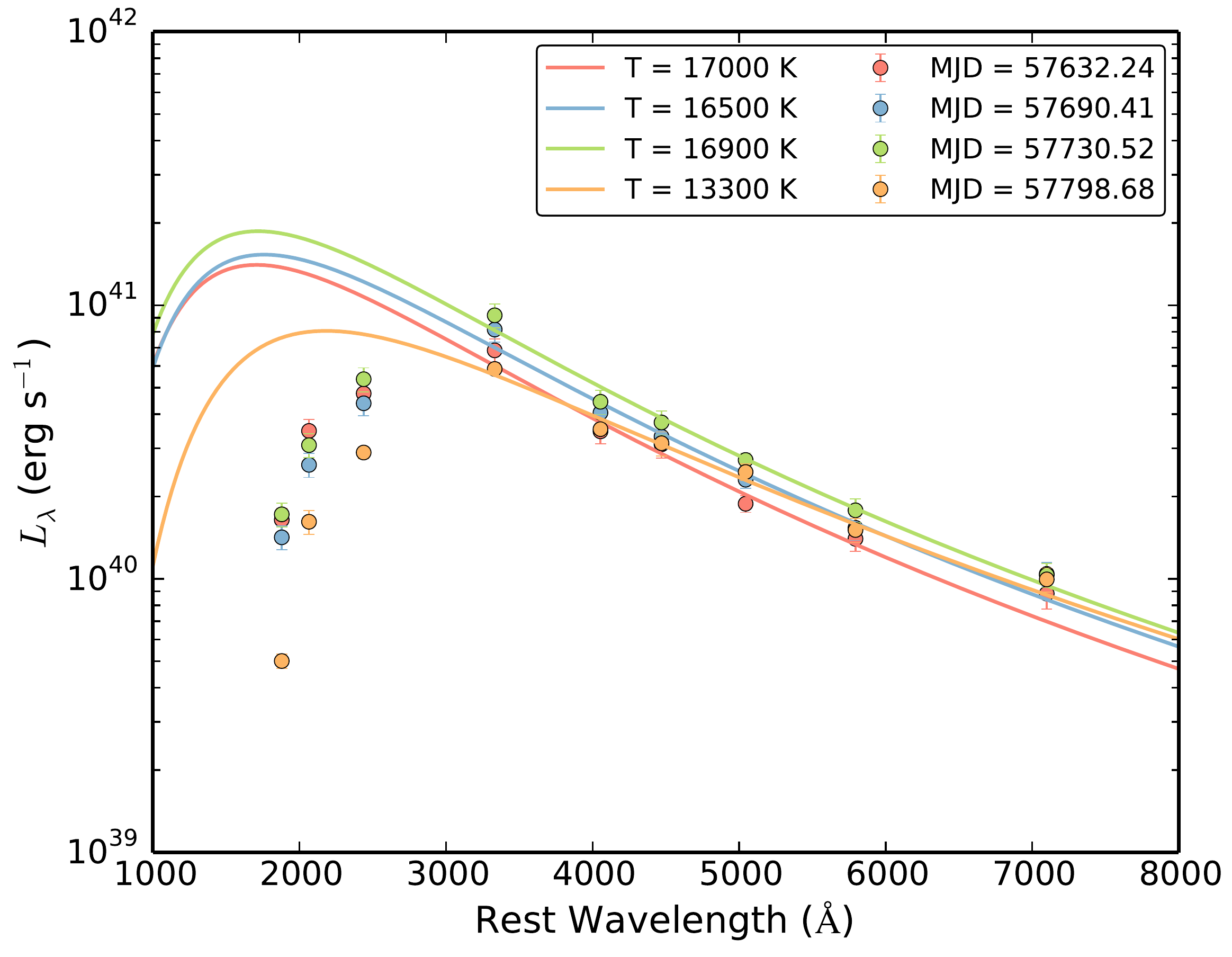}
\end{center}
\caption{UV/optical spectral energy distributions of PS16dtm at three epochs spaced along the plateau phase and one at the final epoch of observations and blackbody fits to the optical data.  The UV flux is significantly suppressed relative to the blackbody curves, indicating the presence of absorbing material.}
\label{SEDs}
\end{figure}

\begin{figure}[t!]
\begin{center}
\includegraphics[scale=0.4]{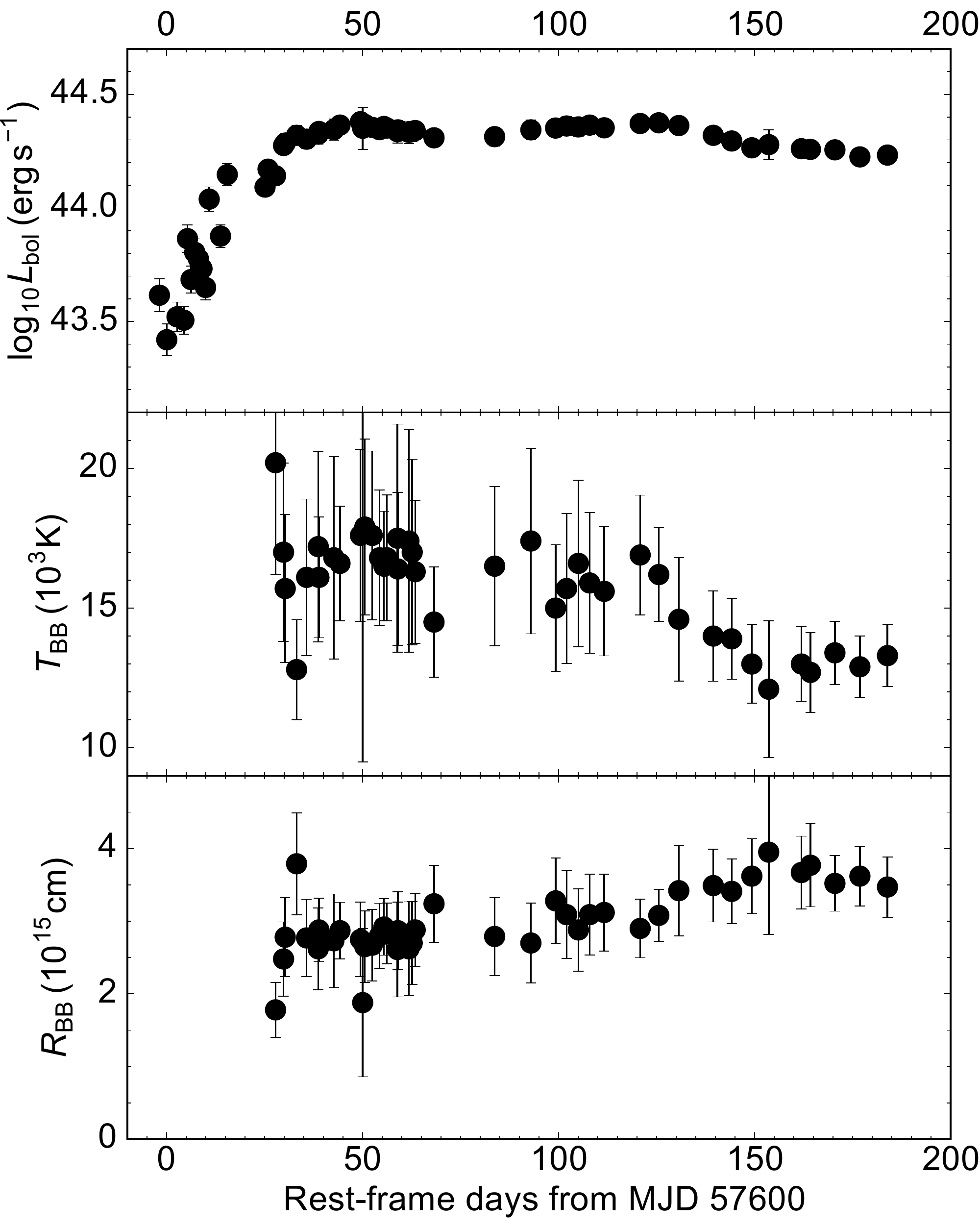}
\end{center}
\caption{\textit{Top:} Rest-frame bolometric light curve of PS16dtm calculated from SEDs constructed using host-subtracted photometry. \textit{Middle}:  Rest-frame blackbody temperature from fits to the optical data only, showing a flat temperature evolution during the plateau phase. \textit{Bottom:} Inferred photospheric radii from the same blackbody fits, calculated assuming a spherical emitting surface.}
\label{bolLC}
\end{figure}

 \begin{figure*}[ht!]
\begin{center}
\includegraphics[scale=0.45]{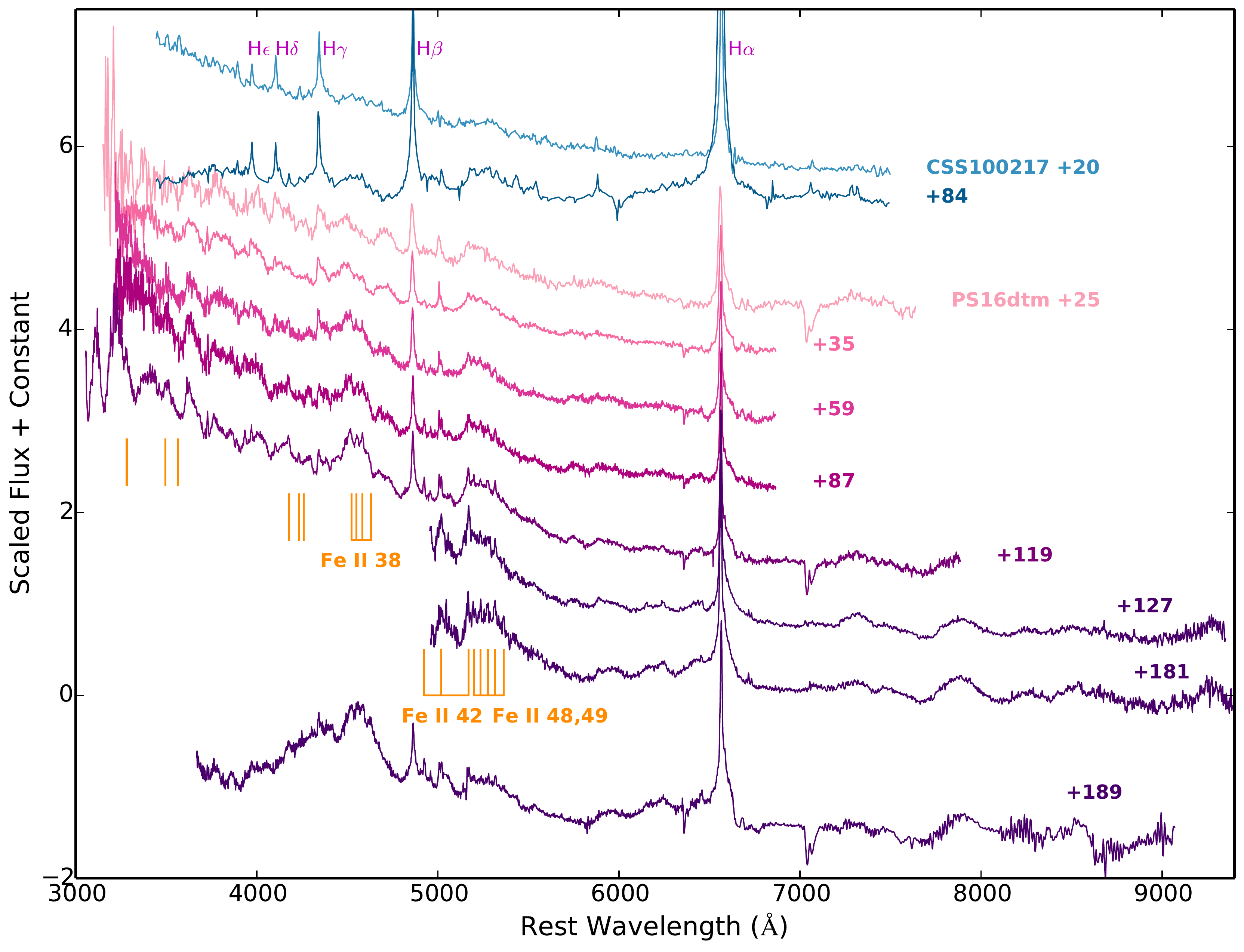}
\end{center}
\caption{Optical spectra of PS16dtm from $+25$ to $+189$ rest-frame days relative to the beginning of the rise.  The $+25$ day spectrum was obtained by \citet{asiagoATel}.  See Table \ref{tab:spec} for details on the spectroscopic observations.  The spectra show very little evolution over $\sim$ 170 days, maintaining a blue continuum with multi-component Balmer and \ion{Fe}{2} emission lines (marked by orange lines).  We also show two epochs of spectra of CSS100217, another luminous transient coincident with the nucleus of a NLS1 galaxy \citep{Drake2011}.}
\label{spec}
\end{figure*}

\section{Observational Characteristics of PS16dtm}
  
\subsection{Multi-Band and Bolometric Light Curves}

In Figure \ref{LC} we show the host-subtracted optical and UV light curves of PS16dtm.  The $V$-band light curve reported by \citet{Dong2016} shows a rise over $\sim 50$ days of about 2 magnitudes above the baseline brightness of the host galaxy.  \citet{Dong2016} note that PS16dtm reached a peak $V$-band magnitude of $\sim 16$, declined by $\sim 0.2$ mag over about 30 days, and then returned to $V \sim$ 16 for about a month.  Our observations show that PS16dtm has continued to remain at $V \sim$ 16 for at least another $\sim$ 80 days with evidence for another modest decline and rebrightening at about 150 days since the beginning of the rise.  In the UV, PS16dtm showed a similar rise (where optical and UV epochs overlap), slight decline, and rebrightening as \citet{DongATel} note, but additional observations show that PS16dtm began to fade in the UV about 100 days after the beginning of the plateau.  PS16dtm has since faded by just over 1 mag in the $uvw2$ filter.        

During the long plateau prior to the onset of UV fading, the observed colors of PS16dtm have remained remarkably constant.  We calculate the $uvw2-u$, $u-v$, $g-r$, and $r-i$ colors and find that they show no trends over this time period with scatter less than 0.1 mag.  This is unusual for SNe, which generally exhibit cooling due to expansion and radiative losses, but similar to TDEs \citep[e.g.,][]{Velzen2011,Gezari2012,Chornock2014}.  The fading in the UV is accompanied by evidence for reddening in the $uvw2 - u$ color, while the optical colors have remained constant.  

Using these multi-band observations we calculate the bolometric luminosity light curve of PS16dtm.  This is accomplished by integrating the observed spectral energy distribution (SED) at each epoch.  We account for the small flux contribution blueward and redward of our photometric coverage by fitting separate blackbodies to the UV and optical data.  A single blackbody provides poor fits due to significant suppression in the UV (see Figure \ref{SEDs}).  In Figure \ref{SEDs} we show example SEDs at four epochs, three during the observed plateau phase and one representing the last epoch, along with blackbody fits to the optical data.  To estimate the bolometric luminosity during the rise phase of the light curve, we extrapolate the multi-band light curves assuming constant colors.  We show the resulting rest-frame bolometric light curve of PS16dtm in Figure \ref{bolLC} along with the temperatures and radii inferred from blackbody fits to the optical data only.  We find that during the plateau phase of PS16dtm, the bolometric luminosity is about $2.2 \times 10^{44}$ ergs s$^{-1}$, which is $\approx 1.8\,L_{\rm Edd}$ for the 10$^{6}$ M$_{\odot}$ black hole at the center of the host galaxy.  This close match is suggestive of an origin for PS16dtm that is related to the supermassive black hole.    

The blackbody fits also indicate a steady temperature of about $1.7\times 10^{4}$ K and a radius of about $3\times 10^{15}$ cm during the plateau phase.  Since PS16dtm began to fade, the temperature has dropped to $1.3\times 10^{4}$ K and the radius has expanded slightly to $3.5\times 10^{15}$ cm.  In addition, the suppressed UV flux blueward of the $U$-band (see Figure \ref{SEDs}) indicates the presence of absorbing gas either within the ejected matter from PS16dtm or local to its environment.  The fading in the UV is likely related to an increase in absorption since the decreasing temperature is not sufficient to explain the reddening of the $uvw2-u$ color.         

In summary, the light curve of PS16dtm is characterized by a rise time of about 50 days, a plateau lasting about 100 days during which it exhibited no color evolution, and a phase dominated by fading in the UV and reddening of the UV colors.  PS16dtm exhibited two episodes of modest dimming and re-brightening at $\sim+75$ and $+150$ rest-frame days since the beginning of the rise.  Moreover, the inferred bolometric luminosity of PS16dtm during the plateau suggests that it radiated at roughly the Eddington luminosity of the supermassive black hole at the center of its host galaxy.

\begin{figure*}[ht!]
\begin{center}
\includegraphics[scale=0.4]{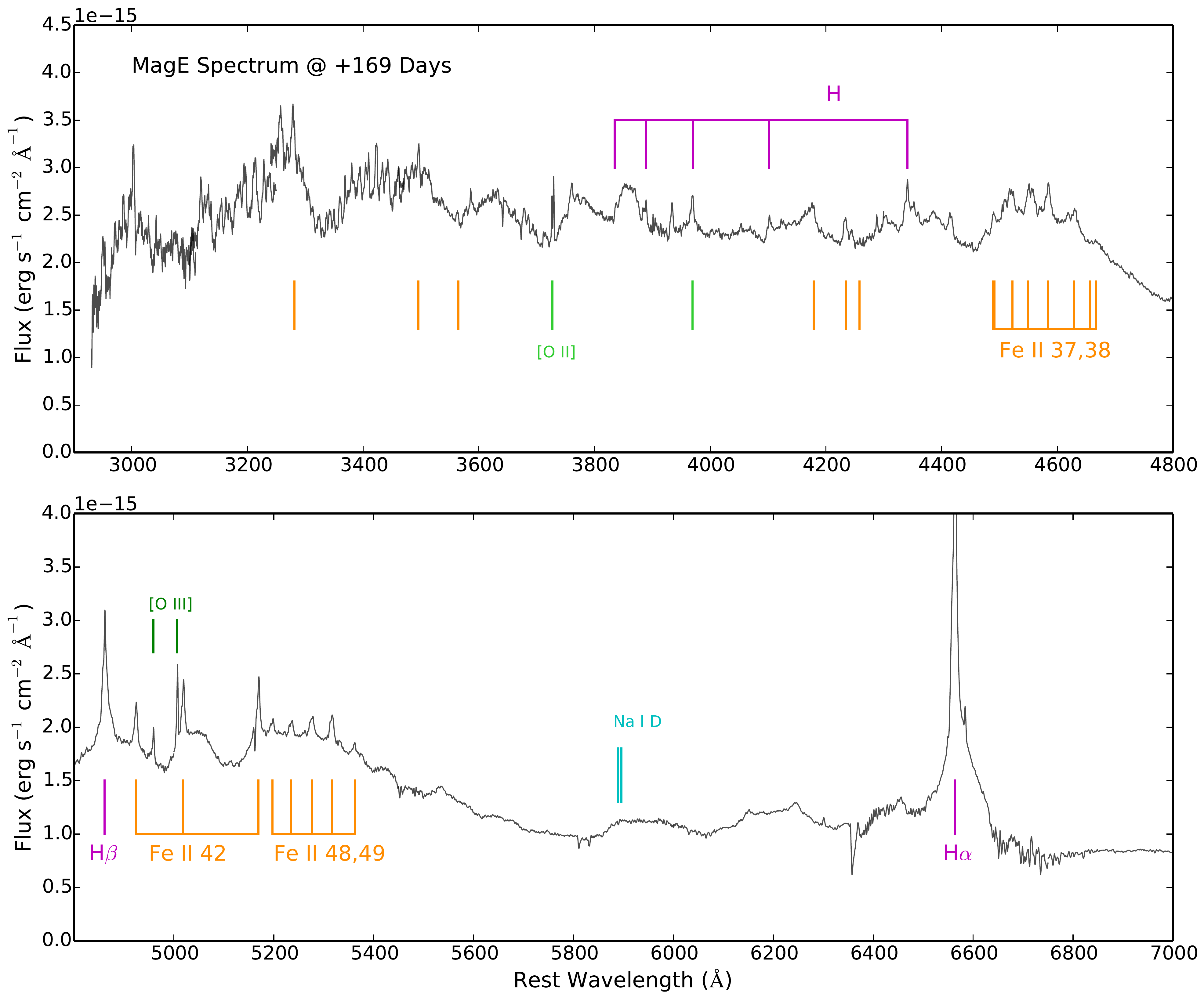}
\end{center}
\caption{Spectrum of PS16dtm obtained using MagE with the Balmer, \ion{Fe}{2}, [\ion{O}{2}], [\ion{O}{3}], and \ion{Na}{1} D lines indicated.  This spectrum resolves the narrow \ion{Fe}{2} lines, which have the same width as the intermediate components of H$\alpha$ and H$\beta$, and clearly illustrates the complex pattern of absorption and emission features in the blue part of the spectrum.}
\label{mage}
\end{figure*}

\begin{figure*}[ht!]
\begin{center}
\subfloat[]{\includegraphics[scale=0.4]{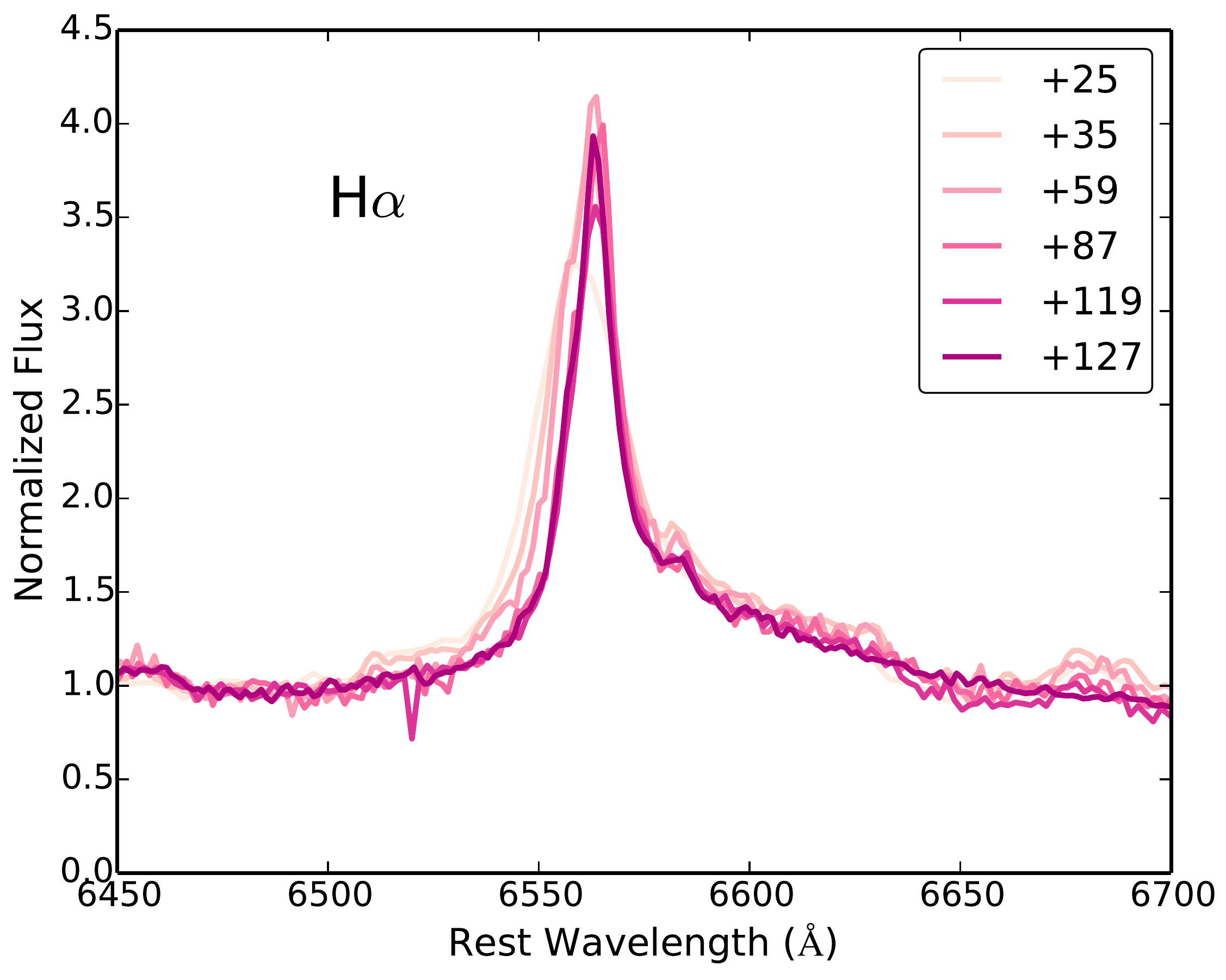}}
\subfloat[]{\includegraphics[scale=0.4]{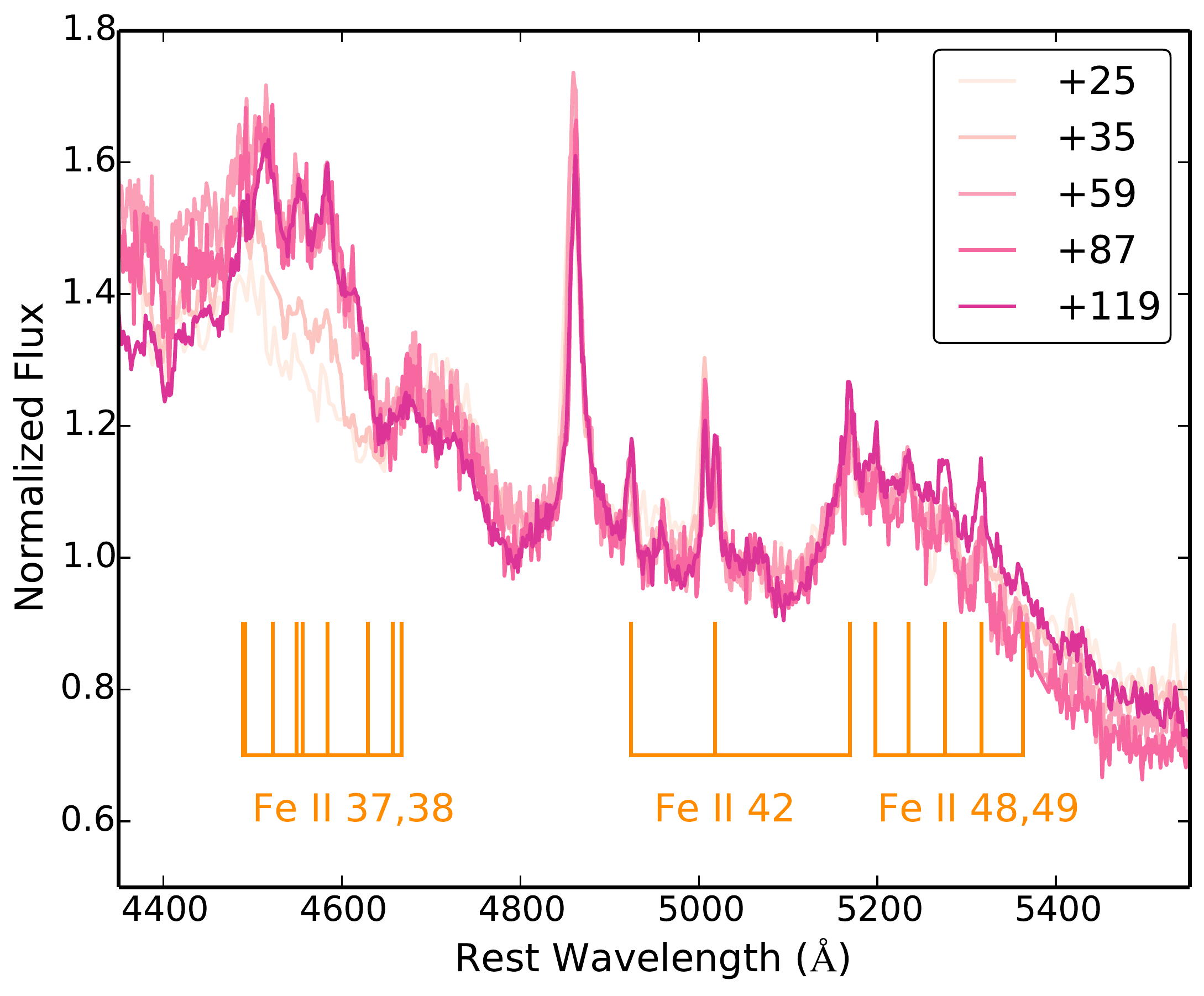}}
\end{center}
\caption{(a) H$\alpha$ profile from our sequence of low-resolution spectra.  The profile is asymmetric with a redshifted broad base, a strong intermediate-width component, and narrow peak. (b) Zoom-in on the region surrounding H$\beta$ where the most prominent \ion{Fe}{2} lines are located that form broad features with superimposed narrow peaks.}
\label{speczoom}
\end{figure*}

\begin{figure*}[ht!]
\begin{center}
\includegraphics[scale=0.4]{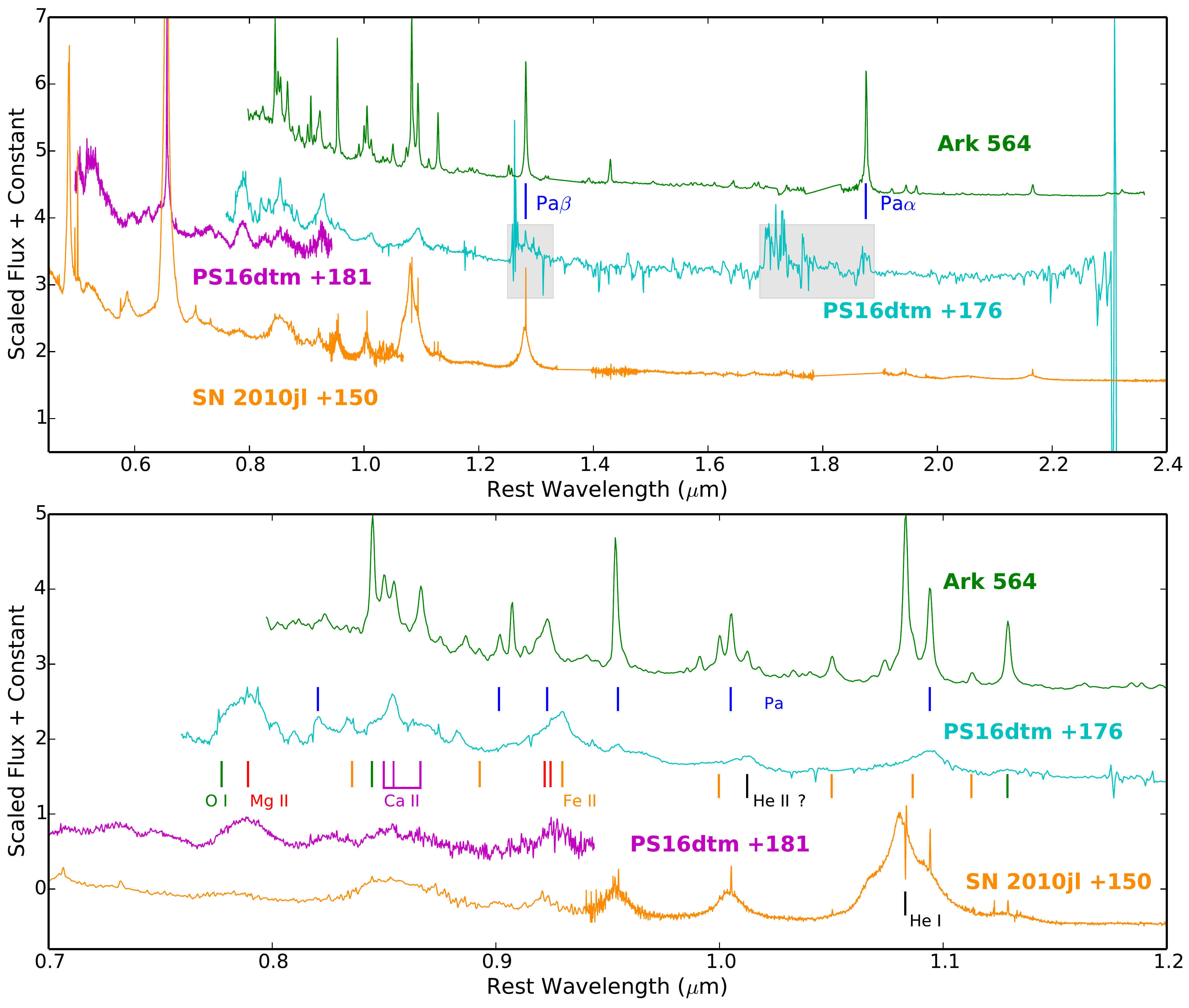}
\end{center}
\caption{NIR spectrum of PS16dtm obtained with FIRE (cyan), the $+181$ day OSMOS spectrum (magenta), a NIR spectrum of NLS1 galaxy Ark 564 \citep[green;][]{Riffel2006}, and a NIR spectrum of SN\,2010jl, a luminous Type IIn SN \citep{Borish2015}.  The shaded regions indicate the location of poorly corrected strong telluric lines.  Lines from the Paschen series, \ion{Fe}{2}, the \ion{Ca}{2} triplet, \ion{Mg}{2}, \ion{He}{2}, and \ion{O}{1} are indicated.  The FIRE spectrum shows several broad asymmetric emission features that may be formed by blends of lines.}
\label{fire}
\end{figure*}
  
\subsection{Spectral Characteristics}
The optical spectroscopic sequence of PS16dtm consisting of the initial classification spectrum \citep{asiagoATel}, our FAST, Blue Channel, OSMOS, and WFCCD observations is shown in Figure \ref{spec} spanning $+25$ to $+189$ rest-frame days since the beginning of the rise (MJD 57600).  We also show the spectrum of CSS100217, another nuclear transient associated with an NLS1 galaxy \citep{Drake2011}.  The higher resolution MagE spectrum of PS16dtm at $+169$ days is shown in Figure \ref{mage}.  The spectra of PS16dtm show remarkably little evolution over the course of $\sim$170 days, consistent with the lack of evolution in the broad-band optical colors.  The spectra show many narrow and broad features superimposed on a blue continuum, including multi-component hydrogen Balmer lines and \ion{Fe}{2} lines.  In the range $\sim 4500 - 5400$ \AA\ there are several complex emission structures with narrow and broad components arising primarily from the \ion{Fe}{2} multiplets 37, 38, 42, 48, and 49.  Whereas in the low resolution spectra, the \ion{Fe}{2} peaks are unresolved, in the MagE spectrum they are resolved into lines that are clearly broader than the lines such as [\ion{O}{2}] and [\ion{O}{3}] that are present in the archival host galaxy spectrum.

Figure \ref{speczoom} shows zoomed in views of H$\alpha$, H$\beta$, and \ion{Fe}{2} lines.  From the $+25$ day spectrum to our $+119$ day spectrum, the \ion{Fe}{2} lines overall become stronger, especially lines belonging to the 37 and 38 multiplets.  Otherwise there are no significant spectral changes in these regions.  H$\alpha$ exhibits a complex asymmetric profile consisting of a broad component and a superimposed narrower component that has a slight shoulder on the blue side of the peak.  The unresolved narrow peak is thus superimposed on a strong intermediate-width component.  While the H$\alpha$ line overall maintains a similar shape in all of our epochs of spectroscopy, the line appears to narrow as a function of time.  The profile of H$\beta$ also displays the same shape.  In addition, the higher resolution of the MagE spectrum revealed that the \ion{Fe}{2} lines also show this same shape, with their broad components adding up to create broad structures with narrower lines superimposed.            

\begin{figure*}[ht!]
\begin{center}
\includegraphics[scale=0.45]{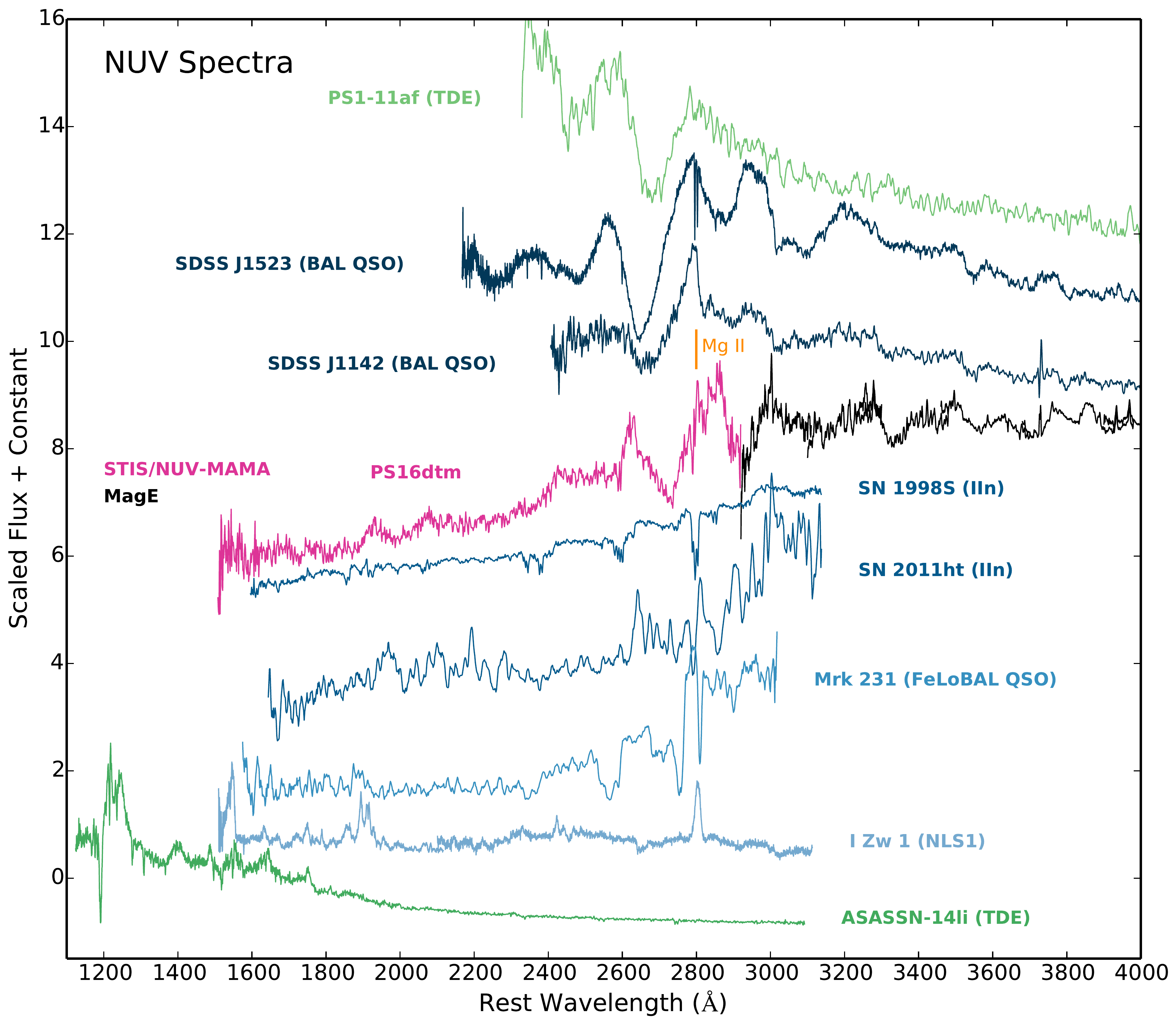}
\end{center}
\caption{NUV spectrum of PS16dtm (magenta) obtained with \textit{HST} using STIS/NUV-MAMA.  Also shown are the blue end of the MagE spectrum of PS16dtm (black) and NUV spectra of several comparison objects including TDEs,  BAL QSOs, Type IIn SNe, and a NLS1 galaxy.  The rest wavelength of \ion{Mg}{2} is marked.  Along with BAL QSOs and PS1-11af, PS16dtm shows broad ($\sim$10000 km s$^{-1}$) blueshifted \ion{Mg}{2} absorption.}
\label{HST}
\end{figure*}

Extending the red coverage with our OSMOS and WFCCD spectra, we detect several broad features near the [\ion{O}{2}] $\lambda\lambda$7320,7330, [\ion{Ca}{2}] $\lambda\lambda 7291,7324$, and \ion{Mg}{2} $\lambda 7892$ lines.  However, given the lack of other broad nebular lines such as [\ion{O}{1}] $\lambda 6300$, we doubt the identification of the bump near 7300 \AA\ as being due to [\ion{O}{2}] or [\ion{Ca}{2}].  In fact, it may be due to a blend of several \ion{Fe}{2} lines.  In Figure \ref{fire} we show the NIR FIRE spectrum of PS16dtm which extends the wavelength coverage to 2.3 $\mu$m.  The spectrum exhibits several broad emission features with asymmetric profiles out to 1.2 $\mu$m, beyond which the spectrum is relatively featureless.  While the presence of strong tellurics near Paschen $\alpha$ and $\beta$ makes their identification difficult, there are peaks near their positions.  Paschen $\gamma$ and other lines in the Paschen series are detected.  Due to the strong presence of \ion{Fe}{2} in the optical spectra, we looked for and identified some NIR \ion{Fe}{2} lines.  Several \ion{Fe}{2} lines are located near the broad asymmetric emission features and might be contributing to creating the asymmetry.  The peak of one of these broad features corresponds with \ion{He}{2} $\lambda 10124$, though the lack of other strong \ion{He}{2} lines (in particular the lack of \ion{He}{2} $\lambda 1640$ in the NUV spectrum; see Figure \ref{HST}) makes this identification uncertain.  The \ion{Ca}{2} triplet shows an unusual shape where the central line in the triplet is much stronger than the others.       
  
Blueward of $\sim 4500$ \AA\ the spectrum shows a very complex combination of narrow features superimposed on broad features.  We identify [\ion{O}{2}] $\lambda\lambda 3727$, Balmer lines, and additional \ion{Fe}{2} lines.  The complexity in this region of the spectrum is likely the result of many multi-component \ion{Fe}{2} lines and transitions of other species common to NLS1 galaxies with similar widths (see comparison with NLS1 spectra in Section \ref{sec:comp}) that cause significant blending.  While many lines appear to be in emission, there is also evidence for absorption, especially the two deep troughs seen at the bluest end of the $+119$ day Blue Channel spectrum.  We also see significant broad absorption in the higher resolution MagE spectrum.  This is consistent with our analysis of the photometric SED, where we inferred that significant absorption must account for the flux suppression in the UV.  The STIS NUV spectrum, shown in Figure \ref{HST}, exhibits a spectral slope consistent with the UV photometry.  The spectrum shows evidence for significant absorption near $\sim$2700 \AA\ which may be blueshifted absorption due to the doublet \ion{Mg}{2} $\lambda\lambda 2800$.  The velocity width of this feature is $\sim 10000$ km s$^{-1}$ and the minimum of the line is blueshifted by $\sim 7000$ km s$^{-1}$ from the rest wavelength of the \ion{Mg}{2} doublet.  There is also an emission line at 2800 \AA\ which is likely \ion{Mg}{2}, with a width similar to the widths of the intermediate components of H$\alpha$, H$\beta$, and the optical \ion{Fe}{2} lines.  Furthermore, some features may be explained by \ion{Fe}{2} emission and absorption as there are many \ion{Fe}{2} resonance lines in the range $\sim 2300 - 2700$ \AA.

\begin{figure*}[ht!]
\begin{center}
\subfloat[]{\includegraphics[scale=0.4]{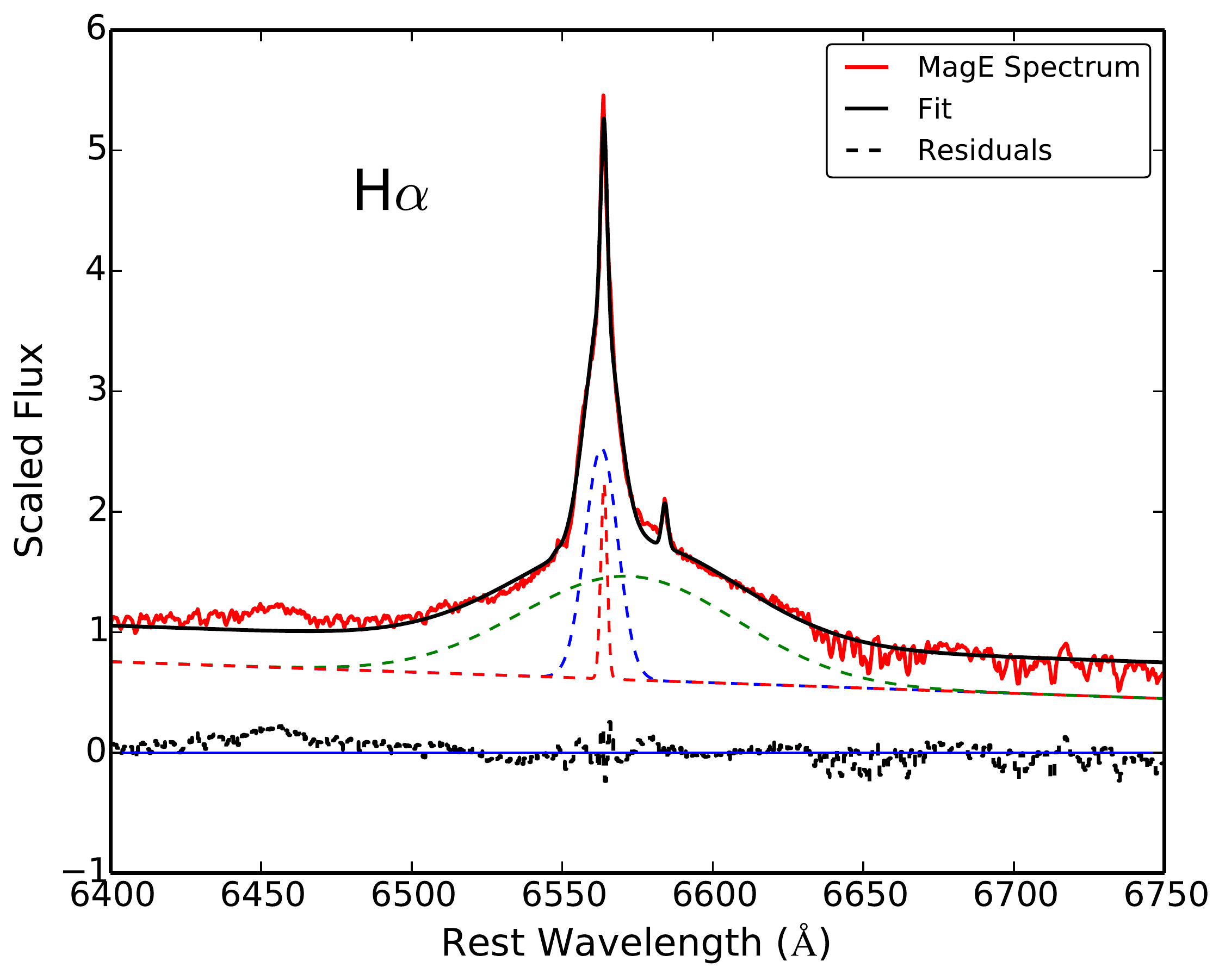}}
\subfloat[]{\includegraphics[scale=0.4]{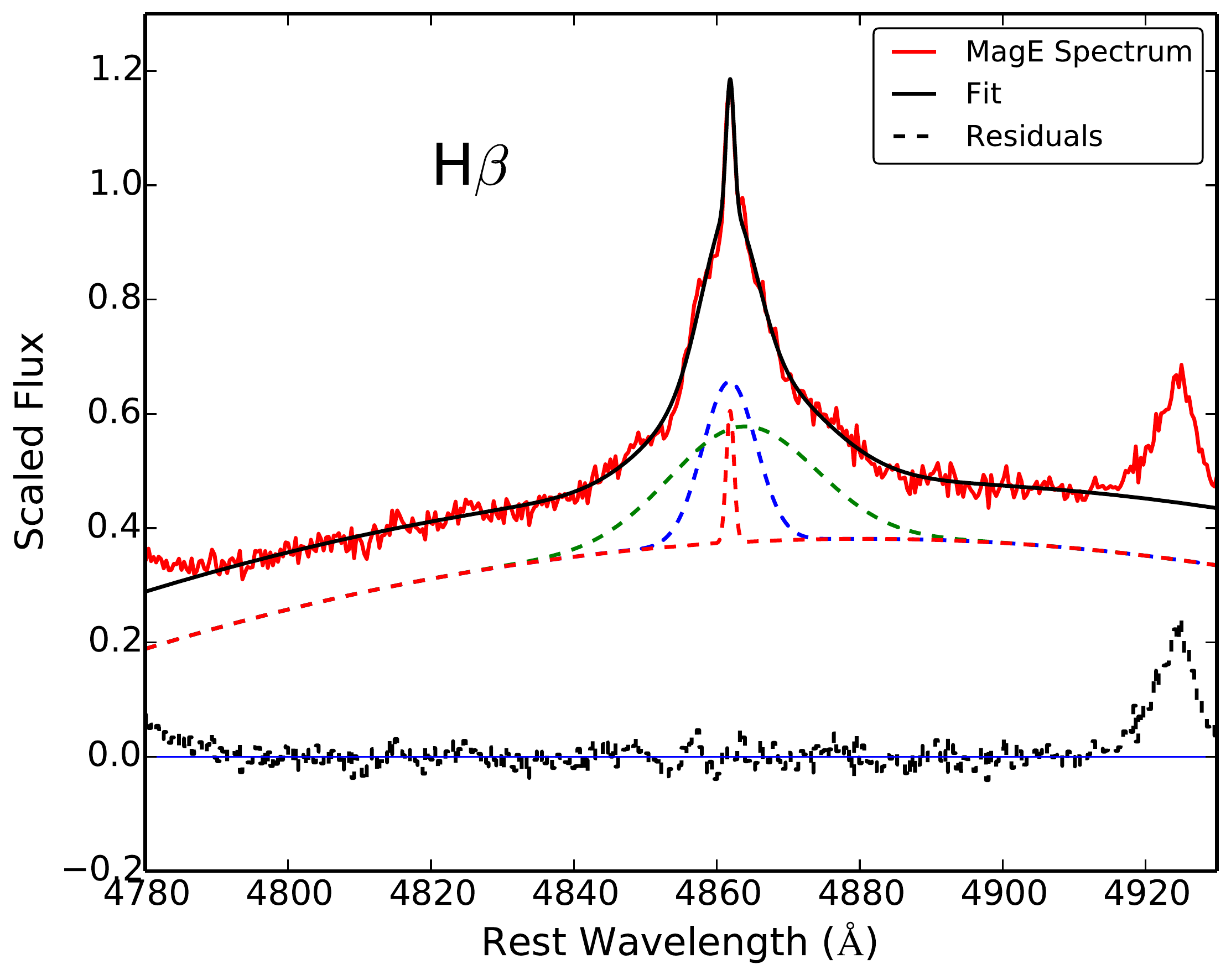}}
\end{center}
\caption{(a) Three-component Gaussian fit (black) to the H$\alpha$ profile from the MagE spectrum (red).  The fit residuals (black dashed) and individual components including the narrow (red dashed), intermediate (blue dashed), and broad (green dashed)  components are also shown. (b) Same as (a) but for the H$\beta$ profile.  This fitting procedure was carried out for each of our spectra.}
\label{magezoom}
\end{figure*}

As shown in Figure \ref{speczoom} and discussed above, the structure of the H$\alpha$ and H$\beta$ profile is asymmetric and has multiple components.  To analyze the complex shape of the H$\alpha$ and H$\beta$ profiles and to compare them with the pre-existing broad component present in the archival host galaxy spectra, we fit them with multi-component Gaussian profiles.  In particular, we model the lines with three Gaussians, one each for the broad, intermediate, and narrow components.  For H$\alpha$ we include the nearby [\ion{N}{2}] lines as they are blended with the broad H$\alpha$ components.  We fix the narrow components at the fixed locations of H$\alpha$ and [\ion{N}{2}] $\lambda$6548,6583 and with widths fixed at the instrumental resolution for our low resolution spectra where the narrow component is unresolved.  We fit for the widths of the narrow components in the higher resolution MagE spectrum.  We simultaneously fit for the parameters of the broad and intermediate components.  We show the combined fits and the individual three components of the H$\alpha$ and H$\beta$ lines in the MagE spectrum in Figure \ref{magezoom}.  

This line fitting analysis shows that the velocity width of the intermediate component of H$\alpha$ at our earliest epochs has not changed significantly relative to the pre-existing component in the SDSS and MagE host spectra, but the equivalent width of this component has increased by a factor of $\sim$2.5. However, this component narrows as a function of time, going from $\sim$900 km s$^{-1}$ in our first epoch to $\sim$600 km s$^{-1}$ in our last. The similarity between the width of our intermediate component and the pre-existing broad component seems to suggest that the component we measure in the transient spectrum may in fact be associated with the broad line region (BLR) of the NLS1 host galaxy, another observation linking PS16dtm to activity near the SMBH.  Moreover, H$\beta$ exhibits the same intermediate component and the MagE spectrum revealed that the \ion{Fe}{2} emission lines also have this component.

In contrast, the flux of the narrow component of H$\alpha$ has remained essentially unchanged relative to the narrow component in the host spectrum. From the MagE spectrum we find that the width of this narrow component is $\sim$100 km s$^{-1}$. The broad component of H$\alpha$, which is not seen in the SDSS host spectrum, has a width of $\sim$3500 km s$^{-1}$ that remains constant during the plateau phase of the light curve and is redshifted by 900 km s$^{-1}$ relative to the narrow component.  During the decline phase of the light curve, the width of this component has increased to $\sim$4000 km s$^{-1}$.          
   
\subsection{X-ray Properties of PS16dtm}
The X-ray non-detection of PS16dtm to a limit 10 times lower than the archival detection suggests that the X-ray emission associated with the AGN has faded or been obscured substantially.  Without knowledge of exactly when the X-rays dimmed, we cannot firmly establish that this dimming is associated with PS16dtm.  However, given that NLS1 galaxies are known to be highly X-ray variable on timescales as short as days \citep{Boller1993,Boller1996,Edelson2002} and that this variability is typically correlated with UV/optical variability \citep{Uttley2006,Grupe2010}, it is significant that a long-term X-ray low state is contemporaneous with a bright UV/optical transient.  Using the X-ray upper limit that we measure ($L_{X} < 1.7 \times 10^{41}$ erg s$^{-1}$) and the optical luminosity estimated from $\lambda L_{\lambda}$(5100\AA) during the plateau phase, the lower limit on the optical to X-ray luminosity ratio is $L_{\rm opt}/L_{X} > 700$.  This signifies a dramatic change from $L_{\rm opt}/L_{X} \sim 0.5$ during the quiescent state.

\section{Comparison to Type IIn SNe, NLS1 Galaxies, and TDEs}
\label{sec:comp}
\begin{figure*}[t!]
\begin{center}
\includegraphics[scale=0.46]{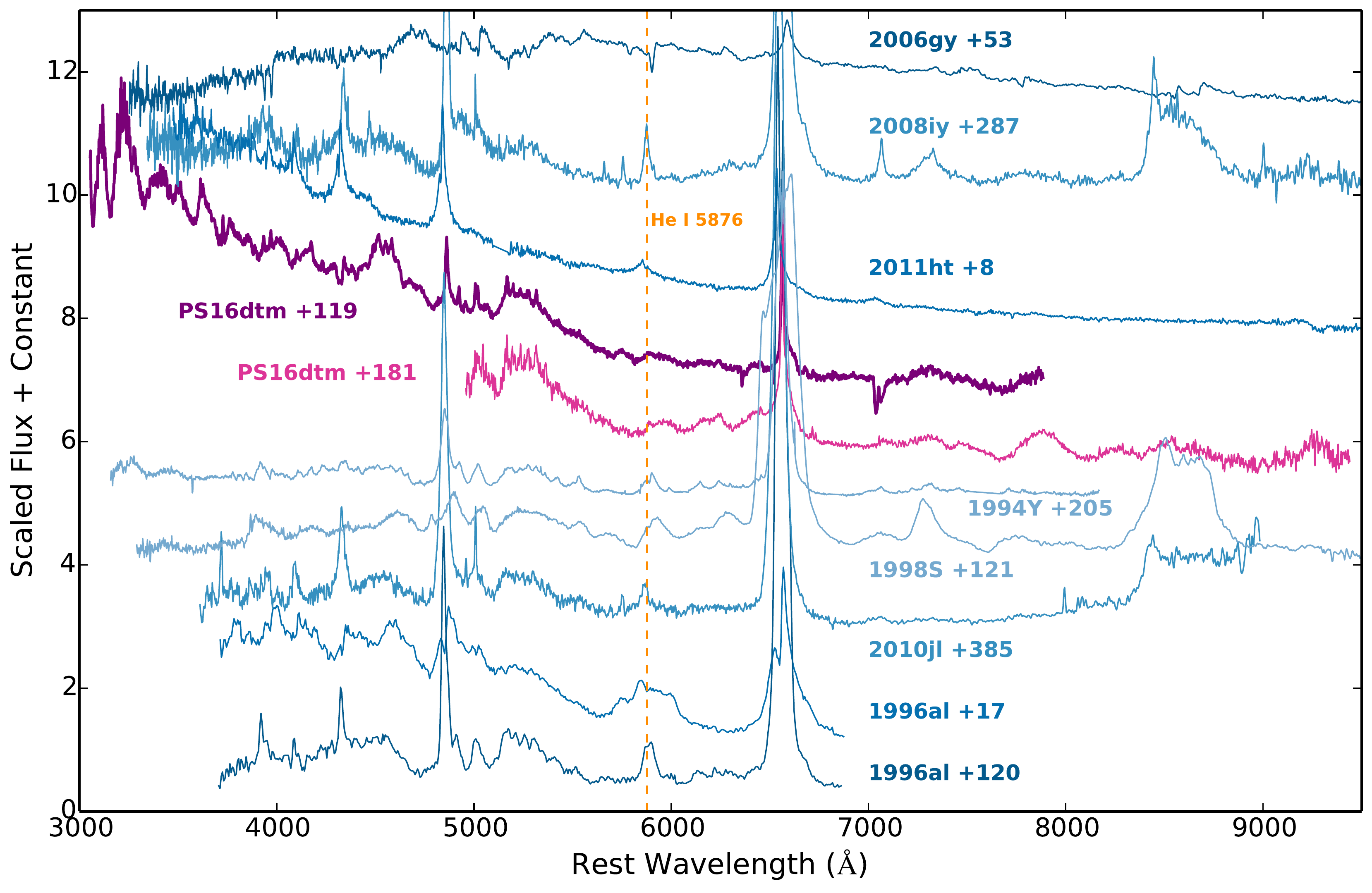}
\end{center}
\caption{Spectroscopic comparison between PS16dtm and several well-studied normal luminosity and superluminous Type IIn SNe.  Our $+119$ and $+181$ day spectra of PS16dtm are shown in magenta and pink, respectively.  While there exist some similarities in the \ion{Fe}{2} features with e.g. SN\,2010jl and SN\,1996al, the overall spectra of typical Type IIn SNe are quite different from PS16dtm.  PS16dtm does not show strong \ion{He}{1} lines as are seen in all of these comparison events, the complexity in the blue part of the spectrum of PS16dtm is not seen in many Type IIn SNe, and the Balmer lines are weaker in PS16dtm.}
\label{compIIn}
\end{figure*}

In Figure \ref{compIIn} we compare the optical spectrum of PS16dtm to the spectra of several well-studied Type IIn SNe: SN\,2006gy \citep{Smith2010}, SN\,2011ht \citep{Humphreys2012}, SN\,2008iy \citep{Miller2010}, SN\,1994Y \citep{Filippenko1997}, SN\,1998S \citep{Leonard2000}, SN\,2010jl \citep{Zhang2012}, and SN\,1996al \citep{Benetti2016}.  We specifically searched for SNe that show spectral similarities to PS16dtm.  We find that some Type IIn SNe show similar \ion{Fe}{2} features to PS16dtm, but typically at $\gtrsim$ 100 days and not as strong relative to the Balmer lines.  Moreover, Type IIn SNe generally exhibit broader and stronger Balmer lines than PS16dtm.  Additionally, PS16dtm displays greater complexity in the blue part of the spectrum than Type IIn SNe, with evidence for absorption in all epochs of PS16dtm's spectra with blue coverage.  Another major distinguishing feature in the comparison sample is the presence of \ion{He}{1}, especially \ion{He}{1} $\lambda$5876, which is not apparent in the spectra of PS16dtm.  Also, the \ion{Ca}{2} NIR triplet is weak relative to what is seen in Type IIn SNe.  Further, the feature at $\sim7890$\AA, which may be \ion{Mg}{2} $\lambda 7892$, is not present in the comparison Type IIn SNe with overlapping wavelength coverage.  In Figure \ref{fire} we show a NIR spectrum of SN\,2010jl \citep{Borish2015}, a luminous Type IIn SN, which exhibits strong Paschen lines and a prominent broad \ion{He}{1} $\lambda 10830$ emission feature.  The NIR spectrum of SN\,2010jl is therefore distinct from that of PS16dtm, which shows no evidence for \ion{He}{1} and generally shows different features.       

\begin{figure*}[t!]
\begin{center}
\includegraphics[scale=0.47]{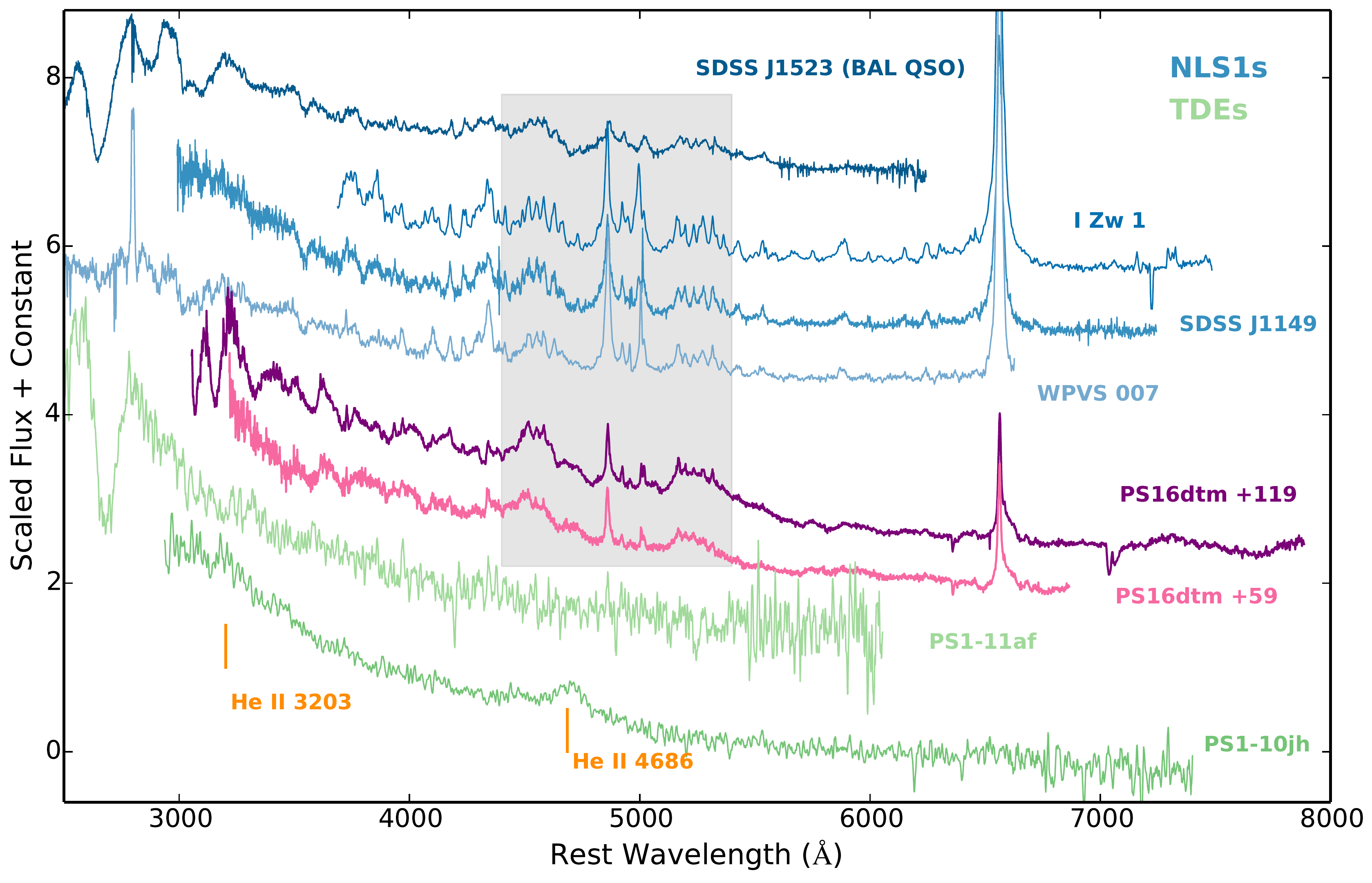}
\end{center}
\caption{Spectroscopic comparison between PS16dtm and three NLS1s, including the well-studied galaxy I Zw 1, one BAL QSO, and two TDEs (PS1-10jh and PS1-11af).  Our $+59$ and $+119$ day spectra of PS16dtm are shown in pink and magenta, respectively.  PS16dtm is strikingly similar to the spectra of NLS1s, most prominently in the \ion{Fe}{2} emission complexes, consistent with a picture in which PS16dtm is an event related to the supermassive black hole and surrounding gas structures.}
\label{compAGN}
\end{figure*}

Given PS16dtm's location at the center of a known NLS1 galaxy, it is natural to compare the spectrum with typical NLS1 spectra to explore whether PS16dtm can be explained as AGN activity.  In Figure \ref{compAGN} we compare PS16dtm to several NLS1 galaxies, including spectra of I Zw 1 \citep{bg92}, WPVS 007 (\textit{HST}/FOS, Proposal 6766, PI: Goodrich), and SDSSJ114954.98+044812.8 \citep{Dong2011}.  I Zw 1 is a well studied NLS1 that has been used to identify and create templates of the \ion{Fe}{2} emission groups.  WPVS 007 is particularly interesting due to its strong X-ray and UV variability \citep{Grupe2013}.  We also show the spectrum of SDSSJ152350.42+391405.2 \citep{Dong2011,Zhang2015}, which is a broad absorption line quasar \citep[BAL QSO;][]{Becker2000}, a class of AGN which show some similarities to NLS1s \citep{Grupe2015}.  PS16dtm is strikingly similar to the spectra of these NLS1 galaxies, particularly in terms of the \ion{Fe}{2} lines.  In addition a few of the many features in the blue part of PS16dtm's spectrum may match some absorption features in the NLS1 spectra, though it appears that PS16dtm displays a more complex spectrum in the region $\sim 3000 - 4200$ \AA\ than the comparison objects.  In Figure \ref{fire} we show a NIR spectrum of NLS1 galaxy Ark 564 \citep{Riffel2006} along with the FIRE spectrum of PS16dtm.  In general, there are greater differences between PS16dtm and this NLS1 galaxy than seen with the optical comparisons.  The emission features in PS16dtm are broader than the lines in Ark 564, though many of the same lines are seen.  

Finally, in Figure \ref{compAGN} we compare the spectrum of PS16dtm to those of two TDEs: PS1-10jh \citep{Gezari2012} and PS1-11af \citep{Chornock2014}.  In general these TDEs exhibit relatively featureless blue continuum emission.  The most notable feature is the broad \ion{He}{2} $\lambda$4686 line seen in the spectrum of PS1-10jh and other TDEs \citep{Arcavi2014}.  As this line is one of the few available for comparison, we have investigated the spectra of PS16dtm to see if \ion{He}{2} is present.  Ultimately, it is difficult to discern whether or not this line is present in our optical spectra due to the presence of nearby \ion{Fe}{2} lines, which create significant blending.  In some TDEs the \ion{He}{2} emission is blueshifted \citep{Arcavi2014}, so if PS16dtm is a TDE with \ion{He}{2} it could be completely blended with the strong \ion{Fe}{2} lines around $4500 - 4600$ \AA.  While the possible identification of \ion{He}{2} in the NIR spectrum suggests that ionized helium may be present, this cannot be confirmed due to the lack of other \ion{He}{2} lines such as the one at 1640 \AA\ that would be expected.  Some TDEs show broad Balmer lines instead of (or in addition to) \ion{He}{2}, with widths ranging from $\sim3400 - 12000$ km s$^{-1}$ \citep{Arcavi2014}, matching the broad component of H$\alpha$ in PS16dtm.  We stress that unlike in TDEs which occur in dormant black holes, the spectrum of PS16dtm reflects the pre-existing NLS1 as well, and is therefore more complex.

In Figure \ref{HST} we show the NUV spectrum of PS16dtm along with several comparison objects including two TDEs \citep[ASASSN-14li, PS1-11af;][]{Cenko2016,Chornock2014}, two BAL QSOs \citep[SDSSJ152350.42+391405.2, SDSSJ114209.01+070957.7;][]{Dong2011,Zhang2015,Liu2015}, an iron low-ionization BAL (FeLoBAL) QSO \citep[Mrk 231;][]{Veilleux2016}, two Type IIn SNe \citep[SN\,1998S, SN\,2011ht;][]{Fransson2005,Humphreys2012}, and the prototypical NLS1 galaxy I Zw 1 \citep{Bechtold2002}.  Many of these objects, including PS16dtm, show a broad ($\sim$10000 km s$^{-1}$) absorption feature near $2600 - 2700$ \AA\ which is often identified as blueshifted \ion{Mg}{2} $\lambda\lambda$2800 absorption due to an outflow.  While the optical spectra of PS16dtm closely resemble NLS1 galaxies, the NUV spectrum exhibits differences with I Zw 1.  Notably, PS16dtm does not show the \ion{C}{4} $\lambda$1540 line that is often strong in NLS1 galaxies, though this may be due to the overall suppression at these wavelengths.  PS16dtm strongly resembles the two BAL QSOs, consistent with a picture in which there is significant absorption from an outflow.  It is likely that similar species identified in the BAL and FeLoBAL QSOs, namely \ion{Mg}{2}, \ion{Mg}{1}, and \ion{Fe}{2}, are responsible for the absorption in PS16dtm.  While PS16dtm is different than the TDE ASASSN-14li in the NUV, PS16dtm and the other TDE, PS1-11af, both show similar \ion{Mg}{2} absorption.  Although the Type IIn SNe NUV spectra have similar spectral slopes to PS16dtm, the Type IIn SNe exhibit a more complex pattern of emission and absorption lines and lack evidence for very broad absorption.

\section{Interpretation of the Properties of PS16dtm}

Before discussing further details about the SN, AGN variability, and TDE interpretations, we summarize the key properties of PS16dtm:

\begin{itemize}
\item The transient is coincident with the nucleus of a NLS1 galaxy that hosts a $\sim 10^{6}$ M$_{\odot}$ black hole. 

\item PS16dtm brightened by about two orders of magnitude relative to the pre-transient AGN luminosity on a timescale of 50 days to a plateau luminosity of $2.2 \times 10^{44}$ ergs s$^{-1}$ for $\sim 100$ days, matching the Eddington luminosity of the SMBH.  The inferred temperature and radius remained roughly constant over this same period.  

\item An X-ray source was detected by \textit{XMM-Newton} at the position of PS16dtm in 2005.  The X-ray emission has since faded to a limit 10 times lower than the detection.  $L_{\rm opt}/L_{\rm X}$ significantly changed from $\sim 0.5$ before the outburst to $> 700$ during the outburst.   

\item The spectra of PS16dtm show little evolution over $\sim170$ days and a striking resemblance to the spectra of NLS1 galaxies.  While exhibiting narrow Balmer lines (which are seen in Type IIn SNe) the spectra are distinct from other Type IIn SNe. 

\item The Balmer and \ion{Fe}{2} lines have multiple components.  The width of the intermediate emission component of H$\alpha$ in the transient spectra is $\sim$750 km s$^{-1}$ averaged over our multiple epochs and narrows from our earliest to latest spectra.  The width in the earliest epoch is similar to the width of the pre-existing broad component in the archival host spectra.  The flux of the narrow component of H$\alpha$ has remained unchanged relative to the host spectrum.  In addition, there is a broad $\sim 3500$ km s$^{-1}$ component that remains constant in width during the plateau. 

\item We see broad ($\sim$10000 km s$^{-1}$) \ion{Mg}{2} absorption in the UV indicative of an outflow.

\end{itemize}

\subsection{Supernova Interpretation}

We have found no SN that matches PS16dtm in most of its spectroscopic features or its evolution.  Most SNe show considerable photometric and spectroscopic evolution due to expansion and cooling, whereas PS16dtm maintains a largely unchanging spectral signature for $\sim 170$ days.  One of the few discernible changes is the strengthening of the \ion{Fe}{2} features known to be associated with the specific type of AGN at the nucleus of PS16dtm's host galaxy.  Furthermore, the AGN contribution to the pre-outburst spectrum is low, so if PS16dtm is a SN one might expect to see more IIn-like features.  Due to the great diversity of Type IIn SNe, the spectral comparison by itself is not sufficient to rule out PS16dtm as a Type IIn SN.  However, when combined with the other properties listed above --- such as its presence at the nucleus of a NLS1 galaxy, unusual light curve, lack of color evolution, and steady high temperature maintained for $\sim$100 days --- it is clear that PS16dtm is not easily explained by the SN scenario.  Most importantly, a SN has difficulty explaining the X-ray dimming and the plateau luminosity matching the Eddington luminosity of the SMBH.  

All lines of evidence point to PS16dtm being related to SMBH activity.  The similarity of the PS16dtm transient spectra to those of NLS1 galaxies, the similarity of the bolometric luminosity to the Eddington luminosity of the SMBH, and the emission line properties indicate that PS16dtm is related to activity near the SMBH.  In addition, the disappearance of the X-ray emission despite a significant UV/optical brightening strengthens the link between PS16dtm and the SMBH.  We therefore consider two remaining possibilities: that PS16dtm can be explained by AGN variability or a TDE.

\subsection{AGN Variability}

AGN have long been known to display variability in the continuum and emission lines \citep[][and references therein]{Peterson2001}, generally thought to be due to changes in the accretion rate possibly related to disk instabilities or thermal fluctuations \citep{Kawaguchi1998,Pereyra2006,Kelly2009}.  Recently, rare modes of variability in multiple types of AGN have been discovered, notably the "changing-look" AGN which show either appearing or disappearing broad emission lines, typically associated with a roughly order of magnitude increase or decrease in the continuum brightness \citep[e.g.][]{Storchi1993,Shappee2014,LaMassa2015,Macleod2016,Gezari2016}.  These changes usually occur on timescales of years, but recently \citet{Gezari2016} reported on iPTF16bco which showed a more rapid $\lesssim$ 1 yr change of a LINER to a broad-line quasar.  Several explanations of these dramatic changes have been proposed, including obscuration by passing dust or gas clouds and accretion rate changes \citep[][and references therein]{Macleod2016}.  

As considered by recent works \citep[e.g.,][]{Shappee2014,Macleod2016}, an accretion rate change scenario could potentially be triggered by viscous perturbations in the disk or an increase in the X-rays originating from the hot corona which irradiate the disk that then reemits at longer wavelengths.  A major issue with the former type of state change is that the timescale for viscous evolution is much longer than the variability timescales observed in changing-look AGN.  The relevant timescale for the latter type of state change is set by the shorter viscous timescales near the inner part of the disk and so may avoid this timescale problem.  \citet{Shappee2014} argue that a time lag between the onset of X-ray and UV/optical variability in NGC 2617 indicates the X-rays are driving the UV/optical variability as in this latter scenario.  Variable obscuration as an explanation for changing-look AGN would require that the UV/optical continuum emitting disk is obscured, cutting off the supply of photons coming to us and the BLR, which may be challenging given the large size of the continuum emitting region.  Invoking timescale arguments, \citet{Macleod2016} and \citet{Gezari2016} argue that the changing-look quasars they studied are better explained by intrinsic accretion state changes rather than variable obscuration.                    

With these recently observed forms of AGN variability in mind, we consider the possibility that PS16dtm represents the outburst state of a low-luminosity AGN whose quiescent state spectrum is dominated by the stellar component of the galaxy (we know the pre-transient AGN contribution is $\sim$5\% of the total light).  When the AGN undergoes some sort of state transition to a much higher accretion rate, the BLR responds to the increased luminosity and the spectrum resembles what we typically see in NLS1s with a stronger AGN component.  Considering that the observed rise time of PS16dtm ($\sim$50 days), which in this scenario is roughly the time for the BLR to respond, is much shorter than the viscous timescale ($\gtrsim 100$ years), it is clear that a state change due to viscous perturbations cannot explain PS16dtm. Furthermore, rapid, two order of magnitude UV/optical flares are not generally seen in AGN (the order of magnitude variability in "changing-look" AGN occur over timescales of years) \citep{Grupe2010,Drake2011}.  Most importantly, this scenario fails to account for the dimming in the X-rays.  As observed in NGC 2617 and iPTF16bco \citep{Shappee2014,Gezari2016}, AGN accretion state changes are expected to show increases in both the X-rays and UV/optical, the opposite of what is observed in PS16dtm. 

The other scenario, involving an increase in X-ray irradiation of the disk (as argued for NGC 2617), must also account for the observed decrease in X-ray emission, since by definition such a flare is caused by an increase in X-rays.  To explain both a decrease in X-rays and increase in UV/optical, this scenario would require a rare accretion rate change that increases the X-rays coupled with obscuration by a dust or gas cloud with sufficient optical depth to obscure the (increased) X-rays to a factor of 10 below the quiescent X-ray level.  Furthermore, one might expect to see time lags from bluer to redder bands as seen in NGC 2617, as the disk is progressively heated, which are not observed in PS16dtm. 

We therefore consider intrinsic AGN variability as an unlikely explanation for the observed behavior of PS16dtm.

\section{PS16dtm as a Tidal Disruption Event}

If the dimming of the X-ray emission is connected to PS16dtm, then obscuration of the inner disk region where the X-rays are produced is required.  An alternative hypothesis for simultaneous obscuration of the X-ray emission and increase in the UV/optical is that the increased accretion giving rise to the transient is the result of a TDE, whose debris stream obscures the X-ray emitting region.  In this scenario, the interaction of the incoming stellar debris with the pre-existing AGN accretion disk or the accretion of the debris stream provides the luminosity to power the rise in the UV/optical continuum and excite the BLR.  Consequently the spectrum we observe looks the same as a NLS1 galaxy with a brighter AGN component.

The TDE hypothesis is the only scenario that can simultaneously reproduce the fast rise and X-ray dimming (difficult to account for in an AGN) and the flat color evolution and unchanging NLS1-like spectrum (showing clear differences with SNe). We therefore explore the TDE interpretation more fully in this section, including a TDE model fit to the bolometric light curve that shows a TDE can also reproduce the light curve shape and luminosity.

\begin{figure*}[t!]
\begin{center}
\includegraphics[scale=0.7]{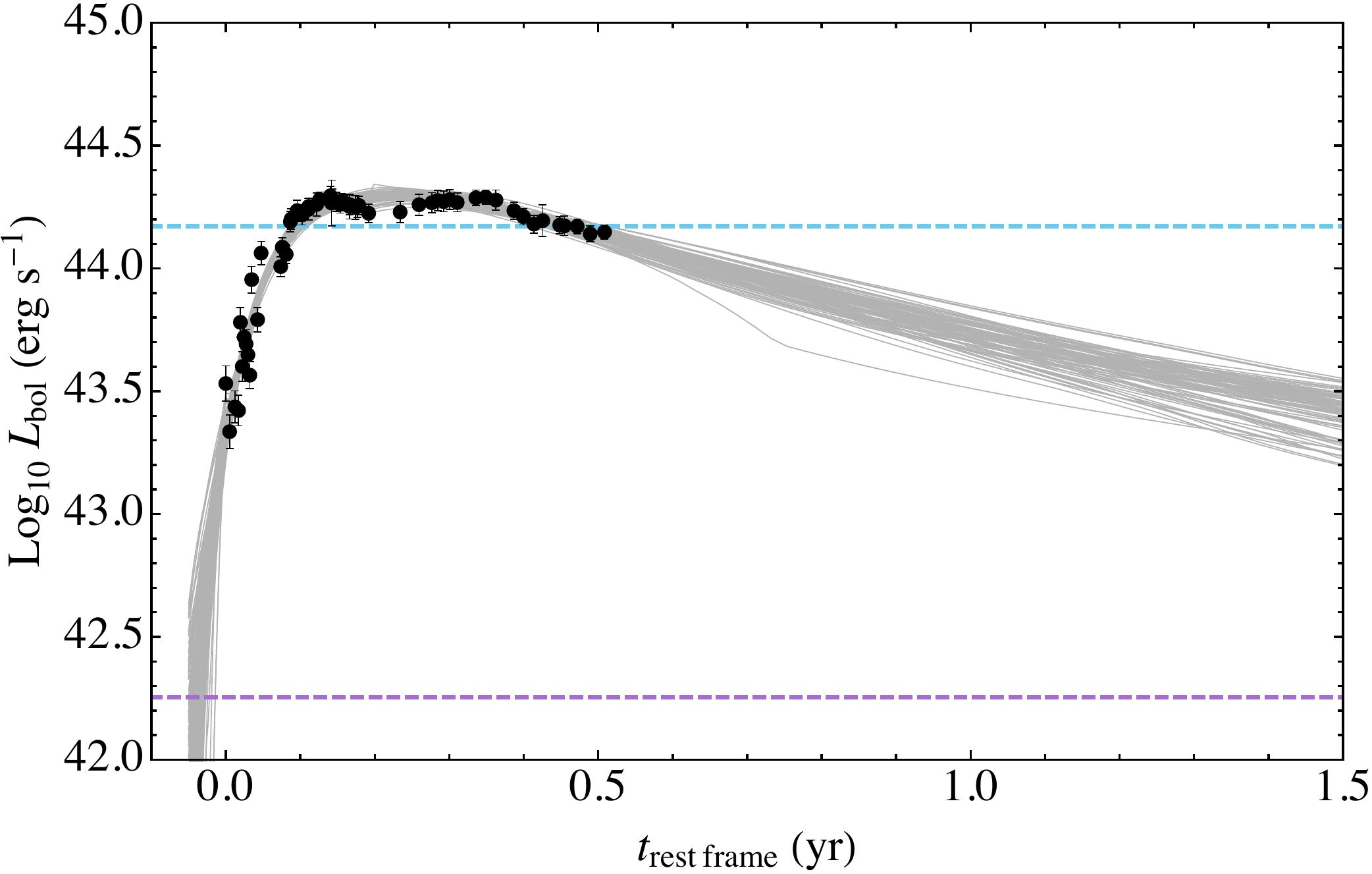}
\end{center}
\caption{Bolometric luminosity of PS16dtm (black points) as compared to an ensemble of model realizations produced with {\tt TDEFit} (gray lines) in the rest frame of PS16dtm. The dashed blue line shows the Eddington luminosity given the assumed fixed black hole mass of $10^6 M_\odot$, whereas the dashed purple line shows the AGN luminosity prior to the flare. The projected behavior of the flare suggests that it will remain bright for years.}
\label{tdefit}
\end{figure*}

\subsection{TDEs in AGN Galaxies}

Because black holes are already luminous in active galaxies, detecting disruptions in them can be challenging, and their variability elicits caution in associating bright flares in such galaxies with TDEs. However, no physical process specifically forbids disruptions from occurring, and in fact tidal disruptions may prefer galaxies with steady accretion onto their black holes \citep{Hills1975}, as the dense star-forming molecular clouds that accompany the accretion can shorten the relaxation time, refilling the loss cone \citep{Perets:2006a}, and also provide a fresh batch of stars to the vicinity of the black hole that may be disrupted.

A significant fraction of supermassive black holes (roughly 10\%) accrete with $L \sim 0.01~L_{\rm Edd}$, comparable to the pre-flare luminosity of PS16dtm. If for the moment we adopt the simple assumption that the flare rate is equal in galaxies of all types, this suggests that approximately 10\% of disruptions should occur in galaxies with $L \gtrsim 0.01~L_{\rm Edd}$. With the number of TDE candidates currently numbering in the dozens \citep{Auchettl:2016a}, the possibility of a tidal disruption having occurred in one or more nearby active galaxies seems likely.

Disruptions about non-accreting black holes with no pre-existing disk may suffer from a ``dark year'' problem where the onset of accretion may be significantly delayed and prolonged, smearing a flare that would usually rise over a period of a month to timescales of up to a decade \citep{Guillochon:2015b}. This issue is especially acute for lower-mass black holes ($M_{\rm h} \lesssim 10^6$), where the fraction of TDEs that are delayed is $\gtrsim 90\%$ due to the debris stream being easily precessed out of its original orbital plane by black hole spin effects \citep{Hayasaki:2016a,Tejeda:2017a}.

If a black hole is active however, the dark year may be avoided by an interaction between the incoming stellar debris and the pre-existing accretion disk. In a naked flare, a stream will continue to elongate and maintain a thin profile until it strikes itself, at which point circularization can begin at the point of collision \citep[e.g.][]{Bonnerot:2016a}. But when the black hole is already accreting, the stellar debris sweeps through the existing accretion disk structure, dissipating its kinetic energy until it runs through a mass comparable to its own. In this interaction between the existing AGN disk and incoming stellar debris, the tangential motion of the disk is more important than that of the stream, as the disk broadsides the stream along a significant fraction of its length.

For an accretion rate onto the black hole $\dot{M}_{\rm AGN}$, the disk structure means that the rate of mass flow in the tangential direction $\dot{M}_\perp$ is larger by a factor $1/(\alpha \mathcal{H}^2)$, where $\alpha$ is the angular momentum transport coefficient ($\sim 10^{-3}$~--~0.1) and $\mathcal{H} \equiv h/r$ ($\sim 10^{-3}$~--~1) is the ratio of the height of the disk to its radius. This means that an incoming debris stream with a fallback rate $\dot{M}_{\rm fb}$ can potentially be halted by an AGN disk even if its radial accretion rate is significantly smaller \citep{Kathirgamaraju:2017a}. When the disruption is in the same plane as the disk, $\dot{M}_{\rm acc} / \dot{M}_{\rm fb} = \alpha \mathcal{H}^2$, which means that assuming typical parameters ($\alpha = 10^{-2}$, $\mathcal{H} = 10^{-2}$), a solar mass star disrupted by a $10^6 M_\odot$ black hole \citep[with peak accretion rate $\sim 100 \dot{M}_{\rm Edd}$,][]{Rees:1988a} can be circularized by a pre-existing accretion disk with an accretion rate of only $\dot{M}_{\rm fb} / \dot{M}_{\rm AGN} = 100 / (\alpha \mathcal{H}^2) = 10^{-4}$.

However, in-plane encounters as described above are extremely rare (only a fraction $\mathcal{H}$ of all encounters), with most disruptions occurring in a plane that is not coincident with the AGN disk plane. For out-of-plane disruptions, the interaction between the stream and the pre-existing disk is much weaker as the stream only intersects a region of the disk with area $h r_{\rm t} \sec \theta$, where $\theta$ is the angle between the accretion disk and the TDE debris stream.

\subsection{Light Curve Fitting}

Fitting the light curve of PS16dtm is complicated by the fact that the flare's spectral energy distribution is contaminated by emission resulting from reprocessing by the broad line region. Additionally, the strong absorption in the UV, which cuts off much more sharply than the usual blackbody cutoff, indicates that the absorption has a strong wavelength dependence, a feature which is characteristic of an optically-thick obscuring layer. This absorption, which depends sensitively on the obscuring layer's density, temperature, and distribution, is likely related to the circularized debris and in the optimal scenario has a simple solution \citep{Loeb:1997a}, but its exact properties are dependent on the parameters of the disruption and can be difficult to ascertain for a given flare \citep{Guillochon:2014a,Roth:2016a}.

With these complications in mind, we fit the light curve of PS16dtm using {\tt TDEFit} \citep{Guillochon:2014a,Vinko:2015a,Alexander:2016a}, but only using its bolometric output rather than its multiband photometry. As in \citet{Alexander:2016a}, we permit the accretion onto the black hole to be viscously delayed with a timescale $\tau_{\rm visc}$. The luminosity from the circularization itself \citep{Piran:2015b,Jiang:2016a} is included in the model, whereas the luminosity of the accretion disk is limited to the Eddington luminosity.

The measurement of the black hole mass from the broad line relation (Section \ref{sec:host}) enables us to fix the black hole mass in our fitting procedure, and thus better constrain the properties of the star within the disruption scenario. Our posterior estimates for the disruption parameters are $\log(M_\ast/\rm{M_\odot}) = -0.66_{-1.21}^{0.17}$, $\beta = 3.0_{-0.5}^{0.5}$ (a full disruption), and $\log(\tau_{\rm visc}/\rm{yrs}) = -0.54_{-1.7}^{0.1}$.  In Figure \ref{tdefit} we show the bolometric light curve of PS16dtm and an ensemble of model realizations from {\tt TDEFit}.  The models are able to reproduce the rise, $\sim$100 day long plateau, and more recent steady decline.  The plateau is likely the result of the viscous delay time, though it may also be the result of the luminosity being capped at the Eddington limit.  However, the short timescale undulations in the light curve are more difficult to account for because the model is not designed to handle the effects due to the likely interaction between the incoming stellar debris and the pre-existing accretion disk.  The models predict that PS16dtm will remain bright for years and therefore continued monitoring of PS16dtm will be an important test of the TDE scenario.

\begin{figure}[t!]
\begin{center}
\includegraphics[scale=0.25]{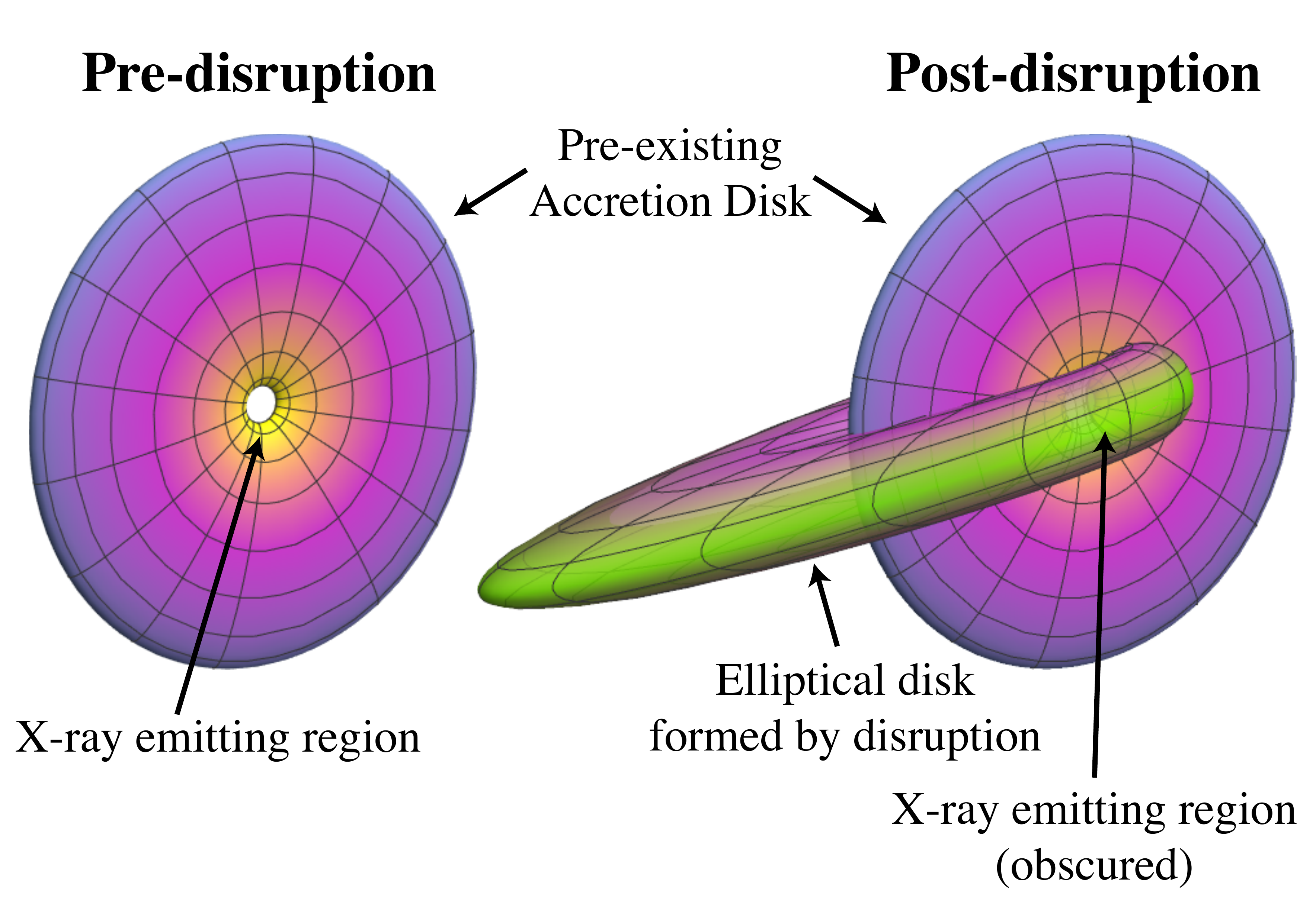}
\end{center}
\caption{An heuristic illustration of how a TDE can obscure the X-ray emitting central region of the pre-existing AGN accretion disk. Prior to disruption, the AGN disk likely exhibited a face-on orientation such that its central regions were visible, whereas the debris disk formed by the disruption is unlikely to have occurred face-on to the observer. Such a structure can also explain why spectral features of the AGN and TDE are simultaneously visible (see Section \ref{specpropTDE}).}
\label{obscure}
\end{figure}

\subsection{Obscuring the AGN's X-ray Emission}
\label{sec:obscure}
In Figure \ref{obscure} we show a heuristic model for how the obscuration of the X-ray emission may be accomplished in the TDE scenario.  The pre-existing accretion disk, from which the X-rays originate near its ISCO, was likely in a face-on orientation to us such that its hot inner regions were within our line of sight. The tidal disruption on the other hand, which does not appear to be emitting its own X-rays, must be in a nearly edge-on configuration.  In this model, the X-ray emitting region near the center of the pre-existing AGN accretion disk is obscured by the accretion disk formed by the stellar debris of the disrupted star.  

If the gas is fully ionized and in a completely spherical configuration, the optical depth for scattering is $270~M_{\rm e, 0.5} r_{i,14}^{-1} r_{o,15}^{-1}$ \citep[definitions within]{Roth:2016a}, which would yield $\tau_{\rm scat} \sim 30$ for the numbers appropriate to PS16dtm. However, the optical depth for absorption alone is $\sim 10^{-4}$ times smaller, with the effective optical depth given by the geometric mean between the scattering and absorption depths, $\tau_{\rm eff} = \sqrt{\tau_{\rm abs} \tau_{\rm scat}} \sim 0.3$.

Three factors are likely to greatly increase this number if the debris structure is edge-on, as we suspect is the case in PS16dtm. First, the finite height of the disk reduces the reprocessing structure's volume by a factor $H/R$, where $H$ is the scale height, increasing $\tau_{\rm eff}$ by the inverse of that factor. Secondly, the debris disk is unlikely to be completely ionized at distances of $\sim 10^{15}$~cm, which yields optical depths up to $10^4$ times larger at 1~keV \citep{Krolik:1984a}. Lastly, X-ray photons that are scattered by the debris disk are more likely to be scattering perpendicular to the disk's midplane, out of our line of sight. These factors combined suggest that $\tau_{\rm eff} \gtrsim 1$ at maximum, and is potentially much larger.

However, the low effective optical depth does suggest that the X-rays may be revealed at a later date in the not too distant future. The total amount of mass in the debris disk once the fallback rate has assumed a power law decay is simply proportional to the fallback rate at that time relative to peak, $\dot{M}/\dot{M}_{\rm peak}$, meaning that $\tau_{\rm eff}$ could easily drop below unity in the next few years as the flare evolves. A definitive upper limit on when the original AGN disk will become visible again is set by when the fallback rate drops below the pre-flare accretion rate, which according to our Monte Carlo ensemble should occur approximately a decade from now.

\subsection{Spectral Properties in the TDE Scenario}
\label{specpropTDE}
In addition to explaining the X-ray obscuration, the debris structure described in Section \ref{sec:obscure} and illustrated in Figure \ref{obscure} also explains how spectral features from the TDE and BLR are simultaneously visible.  The spectra of PS16dtm show a combination of spectral features resulting from the excitation of the broad line region, strong UV absorption, and a $\sim$3500 km s$^{-1}$ emission component associated with H$\alpha$.  Given the likely near face-on orientation of the pre-existing disk, our line of sight will intersect the BLR and therefore we observe emission reprocessed by the BLR.  The broad component of H$\alpha$ corresponds to a radius (assuming Keplerian orbits) near the photospheric radius inferred from the blackbody fits to the photometry.  This suggests that this emission component is coming from the same region that releases the continuum emission of the transient and is generally consistent with a picture in which the luminosity powering the rise in the continuum and the emission lines is coming from the accretion disk or debris stream of the disrupted star.  Furthermore, the strong UV absorption, particularly the broad blueshifted \ion{Mg}{2} absorption, indicates the presence of an outflow that is likely related to the stellar debris.

\section{PS16dtm in a Broader Context}
At least one other well-studied transient occurred at the nucleus of a NLS1 galaxy.  Similar to PS16dtm, the spectra of CSS100217 showed multi-component Balmer lines and strong \ion{Fe}{2}.  Both transients exhibited an additional broad component ($\sim 3500 - 4000$ km s$^{-1}$) not seen in the pre-outburst spectra of their hosts \citep{Drake2011}.  Two transients with generally similar spectra occurring in similar environments is suggestive of a link.  However, there are some differences between the two objects, namely that CSS100217 showed a clear evolution in its continuum (see Figure \ref{spec}) and did not show a plateau in its light curve.  While both objects showed a new broad component, the one in CSS100217 does not remain constant like the component in PS16dtm.    \citet{Drake2011} rule out AGN variability for similar reasons discussed here, primarily that such rapid high-amplitude flares are not seen in the light curves of NLS1 galaxies.  They also argue against a TDE explanation, noting that the temperature and decline rate are inconsistent with theoretical predictions for TDEs.  \citet{Drake2011} conclude that CSS100217 is likely a Type IIn SN based mainly on its photometric, continuum, and emission line evolution.  However, it is now more difficult to argue against a TDE origin for CSS100217 based on its temperature and decline rate, given the diversity of TDEs that have since been observed.  Furthermore, \citet{Drake2011} note that an observed X-ray source associated with CSS100217 was consistent with the soft X-ray emission from TDEs.  Given the lines of evidence suggesting that PS16dtm is a TDE, the similarity between the two events suggest that CSS100217 is also likely a TDE. 

Considering CSS100217 to be a TDE, we can explore whether TDEs occurring in galaxies hosting an AGN have different properties than those occurring in quiescent galaxies.  In addition to PS16dtm and CSS100217, ASASSN-14li may have occurred in an AGN galaxy.  While the X-ray properties of the host galaxy of ASASSN-14li do not indicate an AGN \citep{Miller2015}, there was a pre-existing radio source detected at a luminosity greater than that expected from star formation alone \citep{Alexander:2016a,Holoien2016}.  

\begin{figure}[t!]
\begin{center}
\includegraphics[scale=0.37]{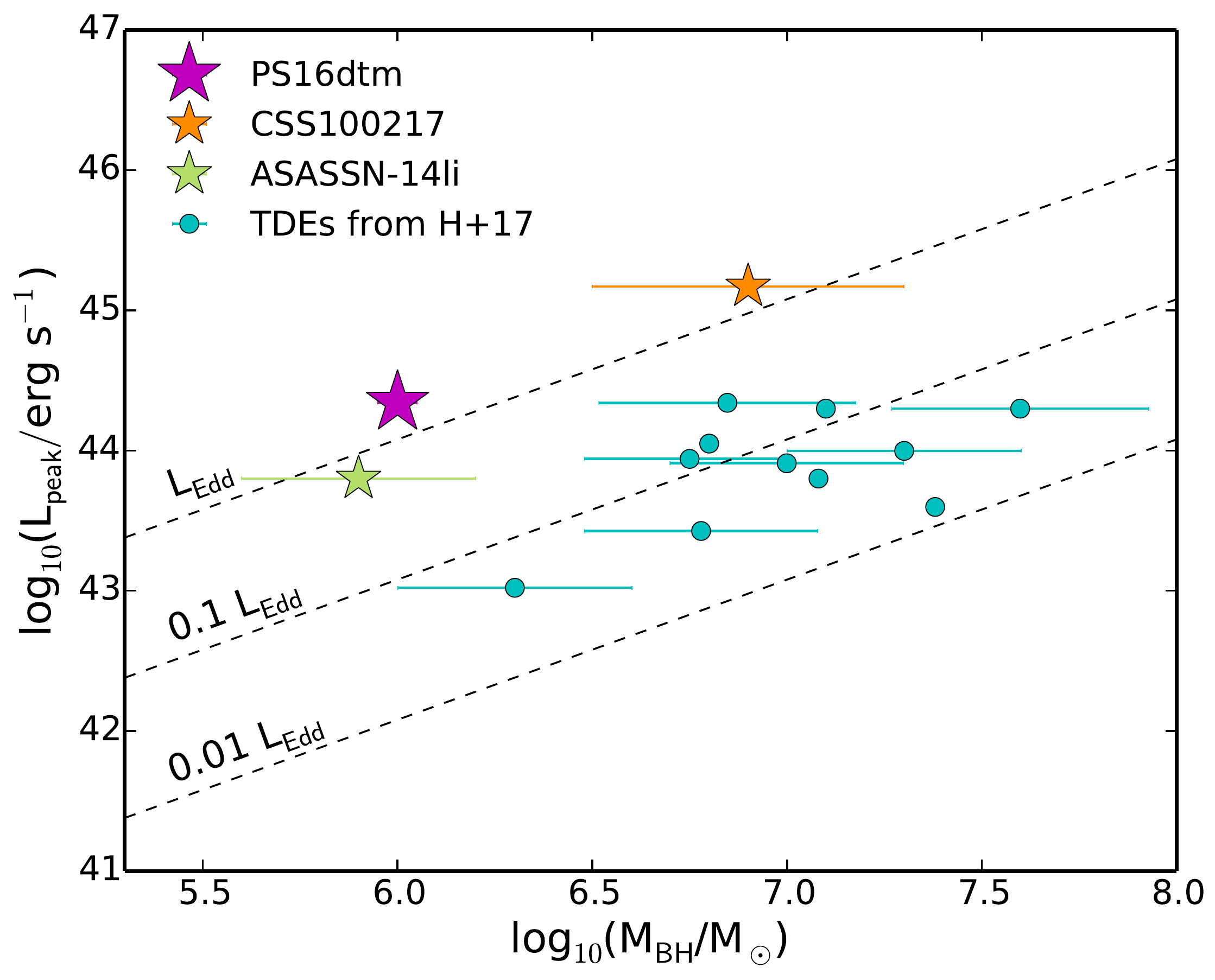}
\end{center}
\caption{Peak luminosity versus black hole mass for PS16dtm, CSS100217, ASASSN-14li, and the sample of TDEs considered by \citet{Hung2017}.  The three dashed lines represent luminosity as a function of black hole mass for Eddington ratios of 1, 0.1, and 0.01.  For ASASSN-14li, we use the black hole mass estimate from \citet{Miller2015}.  This plot suggests the possibility that TDEs which occur in galaxies hosting an AGN exhibit more efficient accretion, allowing them to radiate at the Eddington luminosity of the SMBH, than TDEs which occur in inactive galaxies.}
\label{Lmbh}
\end{figure}

To compare the TDEs which occur in AGN galaxies to those in inactive galaxies, we show in Figure \ref{Lmbh} the peak luminosity ($L_{\rm peak}$) versus black hole mass ($M_{\rm BH}$) for PS16dtm, CSS100217, ASASSN-14li, and the sample of optically discovered TDEs considered by \citet{Hung2017}.  ASASSN-14li was included in the sample of \citet{Hung2017} though here we use a more accurate black hole mass determined by \citet{Miller2015} using {\tt TDEFit} to model the multi-band light curves, which is lower than that used by \citet{Hung2017}.  The remaining TDEs in the sample from \citet{Hung2017} occurred in galaxies without evidence for AGN.  We estimate the mass of the black hole at the center of CSS100217's NLS1 host galaxy using the relation between luminosity and radius of the broad line region \citep{Bentz2013}.  Using the archival SDSS spectrum of the host we measure the width of H$\alpha$ and estimate the AGN continuum luminosity as 50\% of $\lambda L_{\lambda}$(5100\AA).  The uncertainty on the black hole mass is dominated by the uncertainty on the radius of the broad line region since the H$\alpha$ profile shows multiple broad components.  As determined by \citet{Drake2011}, there are three components with velocities of 376, 911, and 2899 km s$^{-1}$.  We therefore take $\sim$900 and 2900 km$^{-1}$ as lower and upper bounds on the width, which translates to the dominant source of uncertainty on the black hole mass, and take the resulting black hole mass to be at the midpoint between the masses set by these two bounds.  

From Figure \ref{Lmbh} it is apparent that PS16dtm, CSS100217, and ASASSN-14li radiated near the Eddington limit of the SMBHs at the center of their hosts, while most of the other optically discovered TDEs radiated at $\sim 0.01 - 0.1\,L_{\rm Edd}$.  This is suggestive of the possibility that TDEs which occur in galaxies hosting an AGN exhibit more efficient accretion than TDEs which occur in inactive galaxies.  This may result from the dissipation of energy due to the interaction of the incoming stellar debris with the pre-existing AGN accretion disk, which may also help with the circularization of the debris.   

\section{Conclusions}
In this paper we present multi-wavelength observations of PS16dtm, a luminous transient coincident with the nucleus of a galaxy hosting an AGN, which was discovered independently by PSST, CRTS, and ASASSN.  Diagnostics from archival data on the host, including emission-line ratios and measurement of a pre-existing broad component to H$\alpha$, suggest that it belongs to the NLS1 class of AGN.  Following discovery PS16dtm rose by two orders of magnitude relative to the pre-transient AGN luminosity over $\sim$50 days to a plateau phase which lasted $\sim$100 days.  Our extensive follow-up and analysis campaign of PS16dtm includes X-ray, UV, optical, NIR, and radio photometry and spectroscopy spanning the rise, plateau, and recent decline phase of the light curve.  We find that the photometric and spectroscopic properties of PS16dtm link the event to SMBH activity.  In particular, the bolometric luminosity during the plateau phase is approximately equal to the Eddington luminosity of the $10^{6}$ M$_{\Sun}$ black hole at the center of the host galaxy.  The optical spectra of PS16dtm closely resemble the spectra of NLS1 galaxies, exhibiting strong multi-component \ion{Fe}{2} and Balmer emission lines.  Perhaps the strongest piece of evidence linking PS16dtm to SMBH activity is the dramatic decrease in the X-ray luminosity during the transient as compared to the archival X-ray detection.   

These observations argue against a SN or AGN variability interpretation.  The unusual light curve shape, lack of color evolution during the long plateau, and spectra are inconsistent with Type IIn SNe.  While AGN have been known to show low level optical variability, the amplitude and timescale of the variability are unusual for AGN, even for the more dramatic changing-look AGN.  Furthermore, a SN or AGN interpretation cannot easily explain the X-ray dimming.  We conclude that PS16dtm is most likely a TDE which provides both the luminosity that powers the rise in the continuum and excitation of the broad line region surrounding the SMBH.  The accretion of the stellar debris obscures the X-ray emitting region near the center of the pre-existing accretion disk, explaining the drop in the X-ray emission.  A TDE model fit to the light curve shows that the disruption of a star with mass $\log(M_\ast/\rm{M_\odot}) = -0.66_{-1.21}^{0.17}$ can explain the shape of the light curve of PS16dtm.  The model predicts that PS16dtm will remain bright for years and we expect the X-rays to reappear once the accretion of the disrupted star is mostly complete.  Comparing TDEs that occurred in galaxies hosting an AGN, like PS16dtm, to those that occurred in normal galaxies, we find that those in AGN may be exhibit more efficient accretion.           

\acknowledgments
The Berger Time-Domain Group at Harvard is supported in part by the NSF under grant AST-1411763 and by NASA under grant NNX15AE50G.  P.K.B. is grateful for support from the National Science Foundation Graduate Research Fellowship Program under Grant No. DGE1144152.  R.C. thanks the Kavli Institute for Theoretical Physics for its hospitality while this work was completed.  This research was supported in part by the National Science Foundation under Grant No. NSF PHY11-25915.  We thank Mike Calkins and Perry Berlind for obtaining the FAST spectra and Allyson Bieryla for assistance in obtaining some of the photometry.  We thank J. A. Klusmeyer for assistance in obtaining the December 2016 OSMOS spectrum.  Based on observations made with the NASA/ESA \textit{Hubble Space Telescope}, obtained from the Data Archive at the Space Telescope Science Institute, which is operated by the Association of Universities for Research in Astronomy, Inc., under NASA contract NAS 5-26555. These observations are associated with program \#14902.  This paper includes data gathered with the 6.5 meter Magellan Telescopes located at Las Campanas Observatory, Chile.  Some observations reported here were obtained at the MMT Observatory, a joint facility of the Smithsonian Institution and the University of Arizona.  This work is based in part on observations obtained at the MDM Observatory, operated by Dartmouth College, Columbia University, Ohio State University, Ohio University, and the University of Michigan.  This paper uses data products produced by the OIR Telescope Data Center, supported by the Smithsonian Astrophysical Observatory.  This paper greatly benefited from the Open Supernova Catalog \citep{Guillochon2017}.

\bibliographystyle{apj}

\begin{thebibliography}{68}

\bibitem[Alexander et al.(2016)]{Alexander:2016a} Alexander, K.~D., Berger, E., Guillochon, J., Zauderer, B.~A., \& Williams, P.~K.~G.\ 2016, \apjl, 819, L25 

\bibitem[Arcavi et al.(2014)]{Arcavi2014} Arcavi, I., Gal-Yam, A., Sullivan, M., et al.\ 2014, \apj, 793, 38 

\bibitem[Auchettl et al.(2016)]{Auchettl:2016a} Auchettl, K., Guillochon, J., \& Ramirez-Ruiz, E.\ 2016, arXiv:1611.02291 

\bibitem[Baldwin et al.(1981)]{BPT} Baldwin, J.~A., Phillips, M.~M., \& Terlevich, R.\ 1981, \pasp, 93, 5

\bibitem[Bechtold et al.(2002)]{Bechtold2002} Bechtold, J., Dobrzycki, A., Wilden, B., et al.\ 2002, \apjs, 140, 143 

\bibitem[Becker et al.(2000)]{Becker2000} Becker, R.~H., White, R.~L., Gregg, M.~D., et al.\ 2000, \apj, 538, 72 

\bibitem[Benetti et al.(2016)]{Benetti2016} Benetti, S., Chugai, N.~N., Utrobin, V.~P., et al.\ 2016, \mnras, 456, 3296 

\bibitem[Bentz et al.(2013)]{Bentz2013} Bentz, M.~C., Denney, K.~D., Grier, C.~J., et al.\ 2013, \apj, 767, 149 

\bibitem[Boller et al.(1993)]{Boller1993} Boller, T., Truemper, J., Molendi, S., et al.\ 1993, \aap, 279, 53 

\bibitem[Boller et al.(1996)]{Boller1996} Boller, T., Brandt, W.~N., \& Fink, H.\ 1996, \aap, 305, 53 

\bibitem[Bonnerot et al.(2016)]{Bonnerot:2016a} Bonnerot, C., Rossi, E.~M., Lodato, G., \& Price, D.~J.\ 2016, \mnras, 455, 2253 

\bibitem[Borish et al.(2015)]{Borish2015} Borish, H.~J., Huang, C., Chevalier, R.~A., et al.\ 2015, \apj, 801, 7

\bibitem[Boroson \& Green(1992)]{bg92} Boroson, T.~A., \& Green, R.~F.\ 1992, \apjs, 80, 109 

\bibitem[\protect\citeauthoryear{{Breeveld} et~al.}{{Breeveld}
  et~al.}{2011}]{Breeveld11}
{Breeveld}, A.~A., {Landsman}, W., {Holland}, S.~T., {Roming}, P., {Kuin},
  N.~P.~M.,  \& {Page}, M.~J. 2011, in American Institute of Physics Conference
  Series, Vol. 1358, American Institute of Physics Conference Series, ed. J.~E.
  {McEnery}, J.~L. {Racusin}, \& N.~{Gehrels}, 373

\bibitem[\protect\citeauthoryear{{Brown} et~al.}{{Brown}
  et~al.}{2009}]{Brown09}
{Brown}, P.~J., et~al. 2009, \aj, 137, 4517

\bibitem[\protect\citeauthoryear{{Burrows} et~al.}{{Burrows}
  et~al.}{2005}]{Burrows05}
{Burrows}, D.~N., et~al. 2005, \ssr, 120, 165

\bibitem[Cenko et al.(2016)]{Cenko2016} Cenko, S.~B., Cucchiara, A., Roth, N., et al.\ 2016, \apjl, 818, L32 

\bibitem[Chomiuk et al.(2011)]{Chomiuk2011} Chomiuk, L., Chornock, R., Soderberg, A.~M., et al.\ 2011, \apj, 743, 114

\bibitem[Chornock et al.(2014)]{Chornock2014} Chornock, R., Berger, E., Gezari, S., et al.\ 2014, \apj, 780, 44

\bibitem[Dong et al.(2011)]{Dong2011} Dong, X.-B., Wang, J.-G., Ho, L.~C., et al.\ 2011, \apj, 736, 86 

\bibitem[Dong et al.(2016)]{Dong2016} Dong, S., Shappee, B.~J., Prieto, J.~L., et al.\ 2016, Science, 351, 257 

\bibitem[Dong et al.(2016)]{DongATel} Dong, S., Chen, P., Bose, S., et al.\ 2016, The Astronomer's Telegram, 9843  

\bibitem[Drake et al.(2009)]{Drake2009} Drake, A.~J., Djorgovski, S.~G., Mahabal, A., et al.\ 2009, \apj, 696, 870

\bibitem[Drake et al.(2011)]{Drake2011} Drake, A.~J., Djorgovski, S.~G., Mahabal, A., et al.\ 2011, \apj, 735, 106

\bibitem[Edelson et al.(2002)]{Edelson2002} Edelson, R., Turner, T.~J., Pounds, K., et al.\ 2002, \apj, 568, 610 

\bibitem[Fabricant et al.(1998)]{Fabricant1998} Fabricant, D., Cheimets, P., Caldwell, N., \& Geary, J.\ 1998, \pasp, 110, 79

\bibitem[Filippenko(1997)]{Filippenko1997} Filippenko, A.~V.\ 1997, \araa, 35, 309

\bibitem[Fransson et al.(2005)]{Fransson2005} Fransson, C., Challis, P.~M., Chevalier, R.~A., et al.\ 2005, \apj, 622, 991 

\bibitem[Gal-Yam(2012)]{GalYam2012} Gal-Yam, A.\ 2012, Science, 337, 927 

\bibitem[\protect\citeauthoryear{{Gehrels} et~al.}{{Gehrels}
  et~al.}{2004}]{Gehrels04}
{Gehrels}, N., et~al. 2004, \apj, 611, 1005

\bibitem[Gezari et al.(2009)]{Gezari2009} Gezari, S., Heckman, T., Cenko, S.~B., et al.\ 2009, \apj, 698, 1367 

\bibitem[Gezari et al.(2012)]{Gezari2012} Gezari, S., Chornock, R., Rest, A., et al.\ 2012, \nat, 485, 217 

\bibitem[Gezari et al.(2016)]{Gezari2016} Gezari, S., Hung, T., Cenko, S.~B., et al.\ 2016, arXiv:1612.04830 

\bibitem[Greene \& Ho(2004)]{gh04} Greene, J.~E., \& Ho, L.~C.\ 2004, \apj, 610, 722

\bibitem[Greene \& Ho(2007)]{gh07} Greene, J.~E., \& Ho, L.~C.\ 2007, \apj, 670, 92 

\bibitem[Grupe et al.(2004)]{gwl+04} Grupe, D., Wills, B.~J., Leighly, K.~M., \& Meusinger, H.\ 2004, \aj, 127, 156

\bibitem[Grupe et al.(2010)]{Grupe2010} Grupe, D., Komossa, S., Leighly, K.~M., \& Page, K.~L.\ 2010, \apjs, 187, 64 

\bibitem[Grupe et al.(2013)]{Grupe2013} Grupe, D., Komossa, S., Scharw{\"a}chter, J., et al.\ 2013, \aj, 146, 78 

\bibitem[Grupe \& Nousek(2015)]{Grupe2015} Grupe, D., \& Nousek, J.~A.\ 2015, \aj, 149, 85 

\bibitem[Guillochon et al.(2014)]{Guillochon:2014a} Guillochon, J., Manukian, H., \& Ramirez-Ruiz, E.\ 2014, \apj, 783, 23 

\bibitem[Guillochon \& Ramirez-Ruiz(2015)]{Guillochon:2015b} Guillochon, J., \& Ramirez-Ruiz, E.\ 2015, \apj, 809, 166 

\bibitem[Guillochon et al.(2017)]{Guillochon2017} Guillochon, J., Parrent, J., Kelley, L.~Z., \& Margutti, R.\ 2017, \apj, 835, 64

\bibitem[Hayasaki et al.(2016)]{Hayasaki:2016a} Hayasaki, K., Stone, N., \& Loeb, A.\ 2016, \mnras, 461, 3760 

\bibitem[Hills(1975)]{Hills1975} Hills, J.~G.\ 1975, \nat, 254, 295 

\bibitem[Holoien et al.(2016)]{Holoien2016} Holoien, T.~W.-S., Kochanek, C.~S., Prieto, J.~L., et al.\ 2016, \mnras, 455, 2918

\bibitem[Humphreys et al.(2012)]{Humphreys2012} Humphreys, R.~M., Davidson, K., Jones, T.~J., et al.\ 2012, \apj, 760, 93 

\bibitem[Huber et al.(2015)]{Huber2015} Huber, M., Chambers, K.~C., Flewelling, H., et al.\ 2015, The Astronomer's Telegram, 7153

\bibitem[Hung et al.(2017)]{Hung2017} Hung, T., Gezari, S., Blagorodnova, N., et al.\ 2017, arXiv:arXiv:1703.01299 

\bibitem[Jiang et al.(2011)]{jgh+11} Jiang, Y.-F., Greene, J.~E., Ho, L.~C., Xiao, T., \& Barth, A.~J.\ 2011, \apj, 742, 68

\bibitem[Jiang (2016)]{Jiang:2016a} Jiang, Y.-F. 2016, In prep.

\bibitem[Kathirgamaraju et al.(2017)]{Kathirgamaraju:2017a} Kathirgamaraju, A., Barniol Duran, R., \& Giannios, D.\ 2017, arXiv:1701.07826 

\bibitem[Kauffmann et al.(2003)]{Kauffmann2003} Kauffmann, G., Heckman, T.~M., Tremonti, C., et al.\ 2003, \mnras, 346, 1055

\bibitem[\protect\citeauthoryear{{Kalberla} et~al.}{{Kalberla}
  et~al.}{2005}]{Kalberla05}
{Kalberla}, P.~M.~W., {Burton}, W.~B., {Hartmann}, D., {Arnal}, E.~M.,
  {Bajaja}, E., {Morras}, R.,  \& {P{\"o}ppel}, W.~G.~L. 2005, \aap, 440, 775

\bibitem[Kawaguchi et al.(1998)]{Kawaguchi1998} Kawaguchi, T., Mineshige, S., Umemura, M., \& Turner, E.~L.\ 1998, \apj, 504, 671

\bibitem[Kelly et al.(2009)]{Kelly2009} Kelly, B.~C., Bechtold, J., \& Siemiginowska, A.\ 2009, \apj, 698, 895-910 
 
\bibitem[Kewley et al.(2001)]{Kewley2001} Kewley, L.~J., Dopita, M.~A., Sutherland, R.~S., Heisler, C.~A., \& Trevena, J.\ 2001, \apj, 556, 121

\bibitem[Kewley et al.(2006)]{Kewley2006} Kewley, L.~J., Groves, B., Kauffmann, G., \& Heckman, T.\ 2006, \mnras, 372, 961 

\bibitem[Komossa(2015)]{Komossa2015} Komossa, S.\ 2015, Journal of High Energy Astrophysics, 7, 148 

\bibitem[Krolik \& Kallman(1984)]{Krolik:1984a} Krolik, J.~H., \& Kallman, T.~R.\ 1984, \apj, 286, 366 

\bibitem[LaMassa et al.(2015)]{LaMassa2015} LaMassa, S.~M., Cales, S., Moran, E.~C., et al.\ 2015, \apj, 800, 144 

\bibitem[Leja et al.(2016)]{ljc+16} Leja, J., Johnson, B.~D., Conroy, C., van Dokkum, P.~G., \& Byler, N.\ 2016, arXiv:1609.09073 

\bibitem[Leloudas et al.(2016)]{Leloudas2016} Leloudas, G., Fraser, M., Stone, N.~C., et al.\ 2016, Nature Astronomy, 1, 0002

\bibitem[Leonard et al.(2000)]{Leonard2000} Leonard, D.~C., Filippenko, A.~V., Barth, A.~J., \& Matheson, T.\ 2000, \apj, 536, 239 

\bibitem[Liu et al.(2015)]{Liu2015} Liu, W.-J., Zhou, H., Ji, T., et al.\ 2015, \apjs, 217, 11 

\bibitem[Loeb \& Ulmer(1997)]{Loeb:1997a} Loeb, A., \& Ulmer, A.\ 1997, \apj, 489, 573 

\bibitem[MacLeod et al.(2016)]{Macleod2016} MacLeod, C.~L., Ross, N.~P., Lawrence, A., et al.\ 2016, \mnras, 457, 389 

\bibitem[\protect\citeauthoryear{{Margutti} et~al.}{{Margutti}
  et~al.}{2013}]{Margutti13}
{Margutti}, R., et~al. 2013, \mnras, 428, 729

\bibitem[Margutti et al.(2016)]{Raf2016} Margutti, R., Metzger, B.~D., Chornock, R., et al.\ 2016, arXiv:1610.01632 

\bibitem[Marshall et al.(2008)]{Marshall2008} Marshall, J.~L., Burles, S., Thompson, I.~B., et al.\ 2008, \procspie, 7014, 701454

\bibitem[Martini et al.(2011)]{Martini2011} Martini, P., Stoll, R., Derwent, M.~A., et al.\ 2011, \pasp, 123, 187

\bibitem[Merloni et al.(2003)]{mhm03} Merloni, A., Heinz, S., \& di Matteo, T.\ 2003, \mnras, 345, 1057

\bibitem[Miller et al.(2010)]{Miller2010} Miller, A.~A., Silverman, J.~M., Butler, N.~R., et al.\ 2010, \mnras, 404, 305

\bibitem[Miller et al.(2015)]{Miller2015} Miller, J.~M., Kaastra, J.~S., Miller, M.~C., et al.\ 2015, \nat, 526, 542

\bibitem[Nicholl et al.(2015)]{Nicholl2015} Nicholl, M., Smartt, S.~J., Jerkstrand, A., et al.\ 2015, \mnras, 452, 3869 

\bibitem[Osterbrock \& Pogge(1985)]{op85} Osterbrock, D.~E., \& Pogge, R.~W.\ 1985, \apj, 297, 166 

\bibitem[Perets et al.(2006)]{Perets:2006a} Perets, H.~B., Hopman, C., \& Alexander, T.\ 2006, Journal of Physics Conference Series, 54, 293 

\bibitem[Pereyra et al.(2006)]{Pereyra2006} Pereyra, N.~A., Vanden Berk, D.~E., Turnshek, D.~A., et al.\ 2006, \apj, 642, 87

\bibitem[Peterson(2001)]{Peterson2001} Peterson, B.~M.\ 2001, Advanced Lectures on the Starburst-AGN, 3 

\bibitem[Piran et al.(2015)]{Piran:2015b} Piran, T., S{\c a}dowski, A., \& Tchekhovskoy, A.\ 2015, \mnras, 453, 157 

\bibitem[Planck Collaboration et 
al.(2014)]{Planck2013} Planck Collaboration, Ade, P.~A.~R., Aghanim, N., et al.\ 2014, \aap, 571, A16

\bibitem[\protect\citeauthoryear{{Pons} \& {Watson}}{{Pons} \&
  {Watson}}{2014}]{Pons14}
{Pons}, E.,  \& {Watson}, M.~G. 2014, \aap, 568, A108

\bibitem[Quimby et al.(2011)]{Quimby2011} Quimby, R.~M., Kulkarni, S.~R., Kasliwal, M.~M., et al.\ 2011, \nat, 474, 487

\bibitem[Rees(1988)]{Rees:1988a} Rees, M.~J.\ 1988, \nat, 333, 523 

\bibitem[Riffel et al.(2006)]{Riffel2006} Riffel, R., Rodr{\'{\i}}guez-Ardila, A., \& Pastoriza, M.~G.\ 2006, \aap, 457, 61 

\bibitem[\protect\citeauthoryear{{Roming} et~al.}{{Roming}
  et~al.}{2005}]{Roming05}
{Roming}, P.~W.~A., et~al. 2005, \ssr, 120, 95

\bibitem[Roth et al.(2016)]{Roth:2016a} Roth, N., Kasen, D., Guillochon, J., \& Ramirez-Ruiz, E.\ 2016, \apj, 827, 3 

\bibitem[Schlafly \& Finkbeiner(2011)]{SF2011} Schlafly, E.~F., \& Finkbeiner, D.~P.\ 2011, \apj, 737, 103

\bibitem[Schmidt et al.(1989)]{Schmidt1989} Schmidt, G.~D., Weymann, R.~J., \& Foltz, C.~B.\ 1989, \pasp, 101, 713 

\bibitem[SDSS Collaboration et al.(2016)]{SDSS2016} SDSS Collaboration, Albareti, F.~D., Allende Prieto, C., et al.\ 2016, arXiv:1608.02013

\bibitem[Shappee et al.(2014)]{Shappee2014} Shappee, B.~J., Prieto, J.~L., Grupe, D., et al.\ 2014, \apj, 788, 48 

\bibitem[Simcoe et al.(2013)]{Simcoe2013} Simcoe, R.~A., Burgasser, A.~J., Schechter, P.~L., et al.\ 2013, \pasp, 125, 270 

\bibitem[Smith et al.(2010)]{Smith2010} Smith, N., Chornock, R., Silverman, J.~M., Filippenko, A.~V., \& Foley, R.~J.\ 2010, \apj, 709, 856 

\bibitem[Smith et al.(2016)]{PSSTATel} Smith, K.~W., Wright, D., Smartt, S.~J., et al.\ 2016, The Astronomer's Telegram, 9401 

\bibitem[Storchi-Bergmann et al.(1993)]{Storchi1993} Storchi-Bergmann, T., Baldwin, J.~A., \& Wilson, A.~S.\ 1993, \apjl, 410, L11 

\bibitem[Tejeda et al.(2017)]{Tejeda:2017a} Tejeda, E., Gafton, E., \& Rosswog, S.\ 2017, arXiv:1701.00303 

\bibitem[Terashima \& Wilson(2003)]{tw03} Terashima, Y., \& Wilson, A.~S.\ 2003, \apj, 583, 145

\bibitem[Terreran et al.(2016)]{asiagoATel} Terreran, G., Berton, M., Benetti, S., et al.\ 2016, The Astronomer's Telegram, 9417

\bibitem[Uttley(2006)]{Uttley2006} Uttley, P.\ 2006, Astronomical Society of the Pacific Conference Series, 360, 101 

\bibitem[van Velzen et al.(2011)]{Velzen2011} van Velzen, S., Farrar, G.~R., Gezari, S., et al.\ 2011, \apj, 741, 73 

\bibitem[Veilleux et al.(2016)]{Veilleux2016} Veilleux, S., Mel{\'e}ndez, M., Tripp, T.~M., Hamann, F., \& Rupke, D.~S.~N.\ 2016, \apj, 825, 42

\bibitem[Vink{\'o} et al.(2015)]{Vinko:2015a} Vink{\'o}, J., Yuan, F., Quimby, R.~M., et al.\ 2015, \apj, 798, 12 

\bibitem[Xiao et al.(2011)]{xbg+11} Xiao, T., Barth, A.~J., Greene, J.~E., et al.\ 2011, \apj, 739, 28 

\bibitem[Zhang et al.(2012)]{Zhang2012} Zhang, T., Wang, X., Wu, C., et al.\ 2012, \aj, 144, 131 

\bibitem[Zhang et al.(2015)]{Zhang2015} Zhang, S., Zhou, H., Wang, T., et al.\ 2015, \apj, 803, 58

\end{thebibliography}

\end{document}